%% file: binary_model_coauthor.tex
\documentclass[12,preprint]{aastex}
\usepackage{graphicx}
\usepackage{natbib}
\usepackage{float}
\usepackage{lscape}

\shorttitle{BINARY CONTAMINATION IN SEGUE} 
\shortauthors{SCHLESINGER ET AL.}

\slugcomment{Submitted to ApJ [29 Jan 2010]}

\newcommand{\msun}{M$_\odot$}

\begin{document}

\title{Binary Contamination in the SEGUE sample: Effects on SSPP 
Determinations of Stellar Atmospheric Parameters }

\author{Katharine J.~Schlesinger\altaffilmark{1},
Jennifer A.~Johnson\altaffilmark{1}, 
Young Sun Lee\altaffilmark{2},
Thomas Masseron\altaffilmark{1},
Brian Yanny\altaffilmark{3},
Constance M.~Rockosi\altaffilmark{4},
B.~Scott Gaudi\altaffilmark{1},
Timothy C. Beers\altaffilmark{2}
}

\altaffiltext{1}{Department of Astronomy, The Ohio State University, 140
W 18th Ave, Columbus, OH 43210} 
\altaffiltext{2}{Department of Physics
and Astronomy and JINA: Joint Institute for Nuclear Astrophysics,
Michigan State University, East Lansing, MI 48824}
\altaffiltext{3}{Fermi National Accelerator Laboratory, P.O. Box 500,
Batavia, IL 60510}
\altaffiltext{4}{UCO/Lick Observatory, University of California, Santa Cruz, CA 95064.}

\begin{abstract}
Using numerical modeling and a grid of synthetic spectra, we examine the
effects that unresolved binaries have on the determination of various
stellar atmospheric parameters for SEGUE targets measured using the
SEGUE Stellar Parameter Pipeline (SSPP).  The SEGUE Survey, a component
of the SDSS-II project focusing on Galactic structure, provides medium
resolution spectroscopy for over 200,000 stars of various spectral types
over a large area on the sky.  To model undetected binaries that may be
in this sample, we use a variety of mass distributions for the primary
and secondary stars in conjunction with empirically determined
relationships for orbital parameters to determine the fraction of G-K
dwarf stars, as defined by SDSS color cuts, that will be blended with a
secondary companion.  We focus on the G-K dwarf sample in SEGUE as it
records the history of chemical enrichment in our galaxy. To determine
the effect of the secondary on the spectroscopic parameters, we
synthesize a grid of model spectra from 3275 to 7850 K ($\sim$0.1 to
1.0\,\msun) and [Fe/H]=$-$0.5 to $-$2.5 from MARCS model atmospheres
using TurboSpectrum. We analyze both ``infinite'' signal-to-noise ratio
($S/N$) models and degraded versions, at median $S/N$ of 50, 25 and 10.
By running individual and combined spectra (representing the binaries)
through the SSPP, we determine that $\sim$10\% of the blended G-K dwarf
pairs with $S/N \geq$25 will have their atmospheric parameter
determinations, in particular temperature and metallicity, noticeably
affected by the presence of an undetected secondary; namely, they will
be shifted beyond the expected SSPP uncertainties. The additional
uncertainty from binarity in targets with $S/N \geq$25 is $\sim$80 K in
temperature and $\sim$0.1 dex in [Fe/H]. As the $S/N$ of targets
decreases, the uncertainties from undetected secondaries increases. For
$S/N$=10, 40\% of the G-K dwarf sample is shifted beyond expected
uncertainties for this $S/N$ in effective temperature and/or
metallicity. To account for the additional uncertainty from binary
contamination at a $S/N \sim$ 10, the most extreme scenario,
uncertainties of $\sim$140 K and $\sim$0.17 dex in [Fe/H] must be added
in quadrature to the published uncertainties of the SSPP.
\end{abstract}


\section{Introduction}
Investigating the chemical evolutionary history of the Milky Way is
critical for understanding galaxy formation and evolution, as we can
accurately measure the abundances of individual stars from rare
populations, such as the most metal-poor stars.  Analyses of cooler
stars, such as G and K dwarfs, are of particular interest, as they have
lifetimes equal to or greater than the age of the Galaxy. The oldest
stars of these types were formed from gas with composition typical of
the earliest moments of galaxy evolution. Simple models of star
formation, such as the closed-box model defined by \citet{schmidt},
reveal the G dwarf problem: we observe fewer metal poor stars than
predicted by simple models of Galactic chemical evolution, indicating
that we do not yet fully understand the early chemical enrichment
history of the Galaxy.  Work such as \citet{pp75}, \citet{norris_ryan},
\citet{sl91}, \citet{wg95}, and later \citet{rpm97}, \citet{chiappini01} and 
\citet{tumlinson06,tumlinson10} investigate this,
giving rise to many different galaxy models with a variety of structural
components and material recycling processes.

One difficulty with these analyses has been the lack of a large uniform
sample of cool stars with accurate metallicity measurements.  Previous
large surveys, such as the Geneva-Copenhagen survey that analyzed
$\sim$14,000 F and G dwarfs in the solar neighborhood, have a large
number of targets but rely on Str{\"o}mgren photometry to determine
stellar metallicity \citep{nordstrom04}.  Similarly, work with SDSS
photometry, such as that by \citet{ivezic08} and \citet{juric08},
utilized a large sample size of stellar targets but determined
metallicities from empirical photometric indicators. Although both of
these analyses boast large sample sizes, they are hindered by the use of
photometric, rather than spectroscopic, metallicity indicators.
Photometric calibrations are susceptible to errors from reddening. They
also have reduced sensitivity for low metallicity targets.  Using
spectroscopy directly can increase the accuracy and precision of
metallicity determinations, in addition to providing kinematic
information such as radial velocities, albeit with the added cost of
increased observing time.

The SEGUE (Sloan Extension for Galactic Understanding and Exploration)
survey combines the extensive uniform data set of photometry from SDSS
with medium resolution (R$\sim$\,1800) spectroscopy over a broad
spectral range (3800-9200\AA) for $\sim$\,240,000 stars over a range of
spectral types \citep{yanny09}.  Technical information about the Sloan
Digital Sky Survey is published on the survey design \citep{york00},
telescope and camera \citep{gunn06, gunn98}, astrometric \citep{pier03}
and photometric accuracy \citep{ivezic04}, photometric system
\citep{fuku96} and calibration \citep{hogg01, smith02, tucker06, pad08}.
Combining SEGUE spectroscopy with SDSS \emph{ugriz} photometry over a
range of 14$<g<$20.3 in $\sim$3500 square degrees on the sky
allows us to better understand the chemical abundance distribution in 
the Galaxy, while avoiding the difficulties associated with purely
photometric surveys and issues of small sample size for spectroscopic
analyses \citep{yanny09,lee08_I}.

SEGUE uses photometric cuts to target SDSS stars for spectroscopic
analysis. The SEGUE ``G dwarf'' sample is defined as having
14.0$<r_0<$20.2 and 0.48$<(g-r)_0<$0.55 while the ``K dwarfs'' have
14.5$<r_0<$19.0 with 0.55$<(g-r)_0<$0.75 where the subscript 0 denotes
dereddened based on the \citet{sfd98} values \citep{yanny09}. This
corresponds to a temperature range of $\approx$5000$-$5300 K for K
dwarfs and $\approx$5300$-$5600 K for G dwarfs for [Fe/H] from $-$0.5 to
$-$2.5.  Each of the spectra is then processed through the SEGUE Stellar
Parameter Pipeline (SSPP) to determine its atmospheric parameters,
namely effective temperature, surface gravity, metallicity, and
$\alpha$-enhancement. The SSPP employs 8 primary methods for the
estimation of T$_{\rm eff}$, 10 for the estimation of $\log g$, and 12 for
the estimation of [Fe/H]. Lastly, the SSPP estimates [$\alpha$/Fe] by
comparison with synthetic spectra utilizing the effective temperature
determined by the SSPP (Lee et al. in prep).  For an in-depth
description of SSPP calculations and processes, see
\citet{lee08_I,lee08_II}. This program's outputs have been checked
against high-resolution spectra of stars within globular and open
clusters, as well as in the field \citep{lee08_II,
allendeprieto08}. Conservatively, the SSPP determines effective
temperatures to within 150 K, surface gravity ($\log g$) to 0.29 dex,
and metallicity ([Fe/H]) to 0.24 dex for targets with
4500$\leq$T$_{eff}$$\leq$7500 with $S/N \geq$50, and can
determine parameters for stars with temperatures as low as 4000 K
\citep{lee08_I}. For spectra with lower signal-to-noise, these
uncertainties increase. For $S/N$=25, $\sigma(T)$=200 K, $\sigma(\log
g)$=0.4 dex, and $\sigma([Fe/H])$=0.3 dex. Lastly, for $S/N$=10,
$\sigma(T)$=260 K, $\sigma(\log g)$=0.6 dex, and $\sigma([Fe/H])$=0.45
dex \citep{lee08_I}. The uncertainties for [$\alpha$/Fe] also increase
with decreasing signal-to-noise. Y.S. Lee et al. (in prep.) show that
errors in [$\alpha$/Fe]$<$0.1 dex can be achieved for spectra with
$S/N>$10. For $S/N$ of or less than 10, $\sigma([\alpha/Fe])\approx$0.2
dex.

It is critical that we understand any potential biases and uncertainties
that arise in this expansive data set, as it will be used extensively to
analyze the structure of the Milky Way. In particular, we focus on
errors associated with unresolved binaries in the SDSS sample. SEGUE
does not have a program designed to check for binaries using repeat
observations. However, there has been work done on potential binaries in
the sample. \citet{sesar2008} have extracted numerous wide field
binaries in SDSS; in particular, they determine the frequency and
distribution of the semimajor axis of these systems.  As they focus on
targets with a separation of greater than 3\arcsec, these targets are
unlikely to affect spectroscopic or photometric parameter
determinations, as each SEGUE spectroscopic fiber is 3\arcsec\,
across. We complement \citet{sesar2008} with an analysis of close
binaries and their effect on atmospheric parameter estimates for SEGUE
targets.

Almost all high-mass stars, such as spectral types O, B, and A, are
likely to be in binaries, while lower mass stars, such as M types, have
a binary fraction of around 30$-$40\%
\citep{fischer92,kouwen08}. Analyses of the F-G stars in the solar
neighborhood have determined that $\sim$65\% of these spectral types
possess at least one companion \citep{dm91}. The effect of the secondary
on the parameter determinations of the primary from photometric and
spectroscopic methods must be quantified and potentially taken into
account when using SEGUE for studies of the Galaxy.  In this analysis,
we first model the binaries in the nearby Galaxy using distributions
based on past empirical studies. We then create a grid of primary and
secondary spectra, which we combine and analyze using the
SSPP. Combining this grid and our modeled distributions, we determine
how the prevalence of binaries will affect the atmospheric parameters
derived by SEGUE. We cover a mass range of primaries from 0.5 to
1.0\,\msun\, and a metallicity range from [Fe/H] of $-$0.5 to
$-$2.5. These techniques can be expanded to different mass and
metallicity ranges.

\section{Pair Modeling}
\label{sec:pair_modeling} 

Every stellar population contains a certain fraction in binaries. The
number of detected pairs depends upon how the sample is observed, such 
as the number and magnitude limits of the observations.  Depending upon
their orbital properties, distances, and mass ratios, some of these
pairs will be blended photometrically, which will potentially change the
measured \emph{ugriz} magnitudes and affect SEGUE target
selection. These targets will also be blended in the spectroscopic data,
affecting the SSPP parameter measurements. For both photometric and
spectroscopic measurements of binaries, the ratio between the members'
luminosities determines the extent of the secondary's effect.

We have used a Monte Carlo simulation to determine the extent of the
influence of undetected binarity. We model a sample of 100,000 binaries,
assigning stellar and orbital parameters based on various empirical
distributions, explained below. For each binary in the sample, we
determine whether or not the pair will appear blended in the data based
upon their orbital properties and distances, namely their period, mass
of each member based on a specific IMF, eccentricity, inclination, and
phase. The point-spread-function (PSF) of the SDSS photometric data is
variable, depending on the seeing. As an approximation to consider the
effects of photometric blending, we set the PSF to 1.4\arcsec, the
median seeing of SDSS in the $r$ band \citep{stoughton02}.
Additionally, if the pair is separated by more than 1.4\arcsec\, but
less than 3\arcsec, they will be spectroscopically blended, i.e. both
stars will be within the SEGUE fiber but the \emph{ugriz} magnitudes
will be separable. All photometrically blended targets are thus
spectroscopically blended, affecting their parameter determinations by
the SSPP. For our purposes, even though they are both spectroscopically
and photometrically blended, we refer to these as photometric blends.
Some pairs will be blended only spectroscopically, affecting SSPP
measurements parameter estimates but leaving their magnitudes
unchanged. We refer to these as spectroscopic blends in the rest of the
paper. Our criteria for spectroscopic blends is conservative, assuming
the largest possible amount of contamination.  When a primary is
centered in a fiber, the contamination from the secondary depends on its
distance to the primary.  By assuming the full contribution from any
secondary closer than 3\arcsec\, we examine the most extreme scenario
possible, and thus determine the upper limit to the spectroscopic
contamination of the sample by undetected secondaries.

While our primary and secondary targets have a range of temperatures and
surface gravities, we assign them the same metallicity, thus neglecting
the effects of any chance superpositions. To determine the effect of
these unassociated pairs on our sample, we use the TriLegal 1.4 program
\citep{girardi05}.  Examining our preliminary G-K dwarf sample from
SEGUE (see \S\,\ref{sec:distance}), we select plates with the highest
number of G-K targets, as these are sky regions of high stellar
density. The magnitude limit for a consistent volume for G and K dwarfs
in our simulation is 17.45 in \emph{r}. However, we also want to
simulate fainter stars that contaminate the sample, similar to our
undetected secondaries. The difference in magnitude between a
0.5\,\msun\, and 0.1\,\msun\, star is around 5 in \emph{r}; thus, our
TriLegal magnitude limit is \emph{r}$\leq$22.45.

We queried TriLegal 1.4 with the coordinates of all the plates with G-K
dwarfs in our preliminary sample, inputting the galactic coordinates for
each plate into TriLegal to determine a star count for a region of 7
deg$^2$, the area on the sky covered by each plate. We then randomly
assigned each star a position within 1$^{\circ}$.49 of the plate
center. Extracting the targets between 0.5\,and 1.0\,\msun, we compare
the coordinates of these to all stars with masses less than that of the
primary, simulating superpositions with an undetected companion within
3\arcsec\, of one another.  The likelihood of a chance superposition in
the entire sample ranges from a minimum of $\sim$2\%, for plate 2313 at
$l \sim$132$^{\circ}$ and $b \sim$$-$63$^{\circ}$, to a maximum of
$\sim$15\%, for plate 1908 with $l \sim$47$^{\circ}$ and $b
\sim$$-$25$^{\circ}$.  For a SEGUE plate directed at the Galactic plane,
the probability of superposition is much higher. The SEGUE plate closest
to the plane of the Galaxy is 2537, at $l \sim$110$^{\circ}$ and $b
\sim$10.50$^{\circ}$ (note that this plate is not in our sample as it
does not have G-K dwarfs due to a different targeting scheme). This
plate has a 44$\%$ likelihood of a superposition for the entire sample,
a very significant contamination. As SEGUE primarily focuses on
latitudes with $|b|>$35$^{\circ}$ \citep{yanny09}, there should
typically be a $\sim$5\% chance of superposition. Simulating these
superpositions would greatly increase the modeling parameter space we
cover, due to the metallicity differences. As there is only a $\sim$5\%
likelihood of encountering one, we do not consider chance superpositions
in our analysis.

We generate a 100,000 pair sample for each combination of primary and
secondary mass distribution at each metallicity from [Fe/H]=$-$0.5 to
$-$2.5, in 0.5 dex increments. An estimated 20,000 stars in the
preliminary SEGUE sample are within the SEGUE G-K color range (see
\S\,\ref{sec:distance}). With our Monte-Carlo simulation of 100,000
pairs, we expect 30,000 to be within the color range, giving us a model
sample slightly larger than the true sample.  To determine the model
uncertainty, which is associated with the finite size of the binary
sample, we use a jackknife error estimate by dividing each 100,000
sample into ten sets of 10,000 pairs and examine the numerical
variation. From this variation, we calculate the standard deviation of
each 10,000 target set from the mean. We then divide by $\sqrt{10}$ to
determine the uncertainty for the larger 100,000 target sample. This is
reported as the uncertainty on the numbers throughout the paper and is
the only source of uncertainty considered.

\subsection{Isochrones}
\label{sec:isochrones}

\begin{figure}
\centering \includegraphics[width=\textwidth]{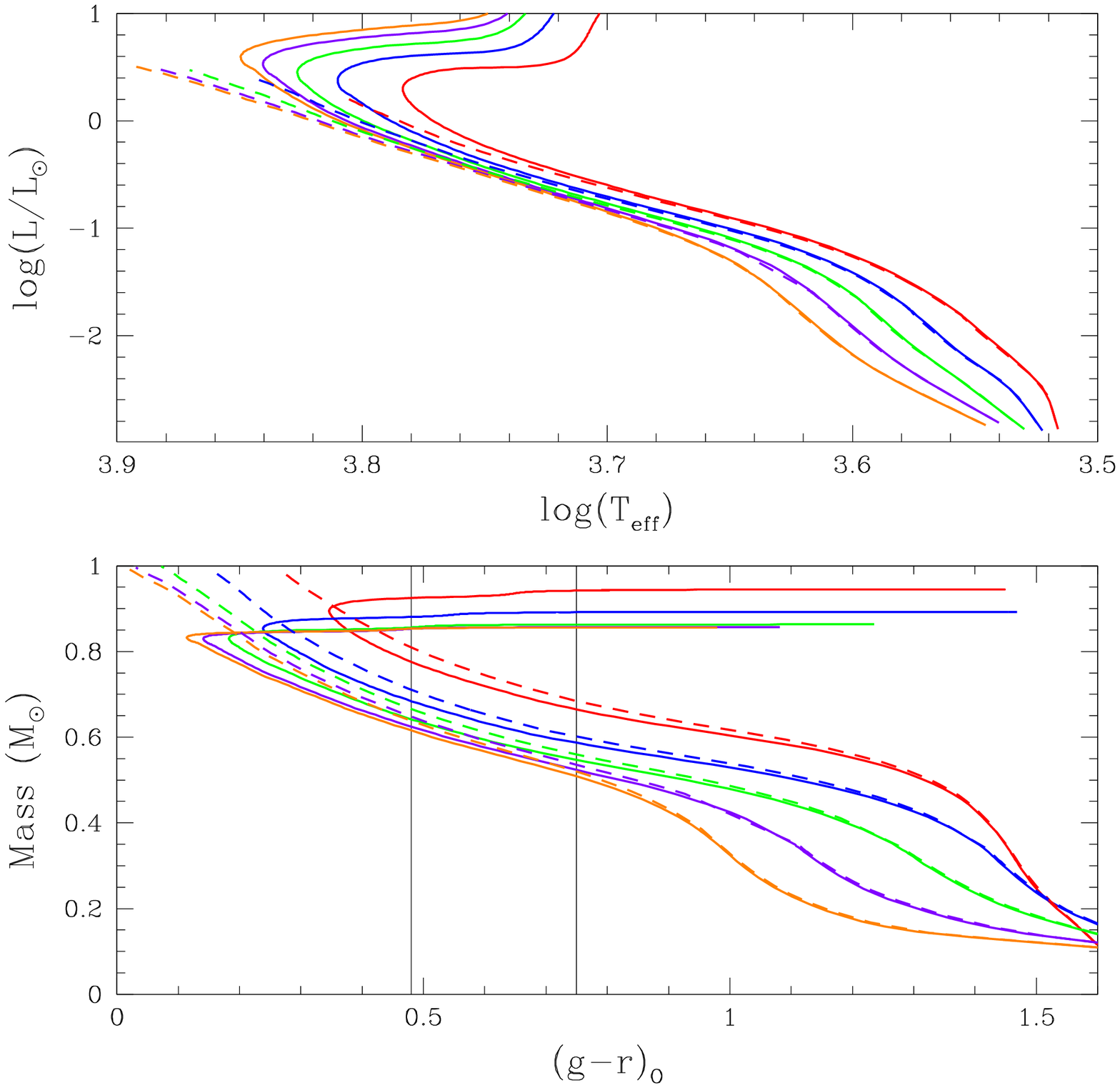}
\figcaption[isochrones3.eps]{Comparison of the Dartmouth isochrones over
the metallicity range of interest for two different ages. Red represents
[Fe/H]=$-$0.5, blue $-$1.0, green $-$1.5, purple $-$2.0 and orange $-$2.5. The
solid lines are 10 Gyr. isochrones and the dashed lines are 3.5 Gyr. For
the main sequence range, the temperature, $(g-r)_0$ color, mass range, and
luminosity discrepancies for the two ages are not large.  Thus, we can
use the younger isochrones for our modeling
purposes. \label{fig:iso_comp}}
\end{figure}

Throughout this analysis, we use isochrones from the Dartmouth Stellar
Evolution database to define temperature, surface gravity, and broadband
$ugriz$ magnitudes for our model stars \citep{dartmouth}.  The
isochrones we select cover a metallicity range from [Fe/H]=$-$0.5 to
$-$2.5, with [$\alpha$/Fe]=0 for masses between 0.1 and
1.0\,\msun\,. This spread reflects the metallicity-mass range of our
preliminary G-K dwarf sample from SEGUE (see \S\,\ref{sec:distance}).
We examine only the main sequence section of the selected
isochrones. Additionally, our model atmospheres limit us to stars cooler
than 8000 K. Although we expect the SEGUE stars to be around 10 Gyr in
age, to simulate targets on the main sequence up to 8000 K requires
using a 3.5 Gyr isochrone (see Fig.\ref{fig:iso_comp}).  Even though
this age is less than the expected age for these stars, there are not
significant differences between the isochrones of various ages on the
main sequence (see Fig.\ref{fig:iso_comp}).  The Dartmouth isochrones
are especially useful as they are calculated directly for $ugriz$
photometry and do not require any conversions between $UBVRI$ and SDSS
photometry, which could be a potential source of uncertainty as these
transformations are metallicity dependent and incomplete over our
temperature and metallicity ranges. We also used the isochrones from the
Padova database \citep{girardi}, which cover a similar metallicity and
age range. Using these two sets of isochrones provided similar numerical
results, and we selected the Dartmouth isochrones for the bulk of our
work as they are in uniform incremental steps of metallicity, making our
distribution of models over metallicity smoother.

\subsection{Binary Properties}
\label{sec:binary properties}
Previous analyses of G dwarf binaries in the solar neighborhood, in
particular \citet{dm91}, have established empirical expressions for
various orbital properties, such as period and eccentricity. By
utilizing these, in conjunction with different empirical descriptions of
stellar mass functions, we can model a sample of binaries that mimic
observed conditions. Our adopted model parameters for the synthetic
pairs follows.

\subsubsection{Period}
We assign each pair a period between 0 and 10$^{10}$ days based on the
Gaussian distribution in $\log P$ from \citet{dm91}. This work analyzed
the properties of 82 pairs with G dwarf primaries in the solar
neighborhood from the CORAVEL and Gliese catalogs, empirically fitting a
lognormal to the period distribution with an average $\log P= $4.8 and
$\sigma_{\log P}= $2.3, where P is in days.

\subsubsection{Primary Mass Distribution} 
We use three different distributions to define the masses of the
primaries.  The color range of our SEGUE G-K dwarf sample is 
0.48$\leq$$(g-r)_0$$\leq$0.75, implying a mass range from approximately
0.5 to 0.8\,\msun\, using the Dartmouth isochrones over a range of
metallicities (see Fig.\,\ref{fig:iso_comp}). Our mass range extends
slightly beyond the typical mass range of G and K dwarfs, going from 0.5
to 1.0\,\msun\,, to ensure that we account for all potential
contaminants; higher mass primaries can be bumped into the color region
of G-K dwarfs when blended with cooler secondaries. Similarly, G-K dwarf
primaries can potentially be bumped out of a rigid color cut when
blended with a secondary.

We use three models for the mass distribution of the primary stars from
\citet{salpeter55}, \citet{kroupa01}, and \citet{chabrier03}. The
Salpeter distribution, based upon the observed luminosity function at
that time, is a simple power law:
\begin{equation}
f(M)\propto M^{-2.35}
\end{equation}
The \citet{kroupa01} model is in the form of three broken power laws
over different mass ranges and is quite similar to a Salpeter
distribution over the mass range of the primaries.  For the primaries,
\begin{equation}
f(M)\propto M^{-2.3}
\end{equation}

Lastly, work by \citet{chabrier03} utilized observational data of the
bottom of the main sequence to determine that the basic power law
relationships defined by Kroupa and Salpeter are not accurate for stars
below \msun.  The Chabrier distribution for stars with
M$<$\msun\, is based on a Gaussian form:
\begin{equation}
f(M)\propto \exp[-(\log M -\log m_{c})/2\sigma^{2}]
\end{equation}
where $m_{c}$ is the mean mass, 0.079 \msun\,, and $\sigma$ is the variance in
$\log (M/$\msun\,$)$, 0.69. Chabrier also determines a mass function for system
masses to account for unresolved binaries. This function is consistent
with the single mass function with a 50\% binary fraction
\citep{chabrier03}. As we want the individual mass of primary and
secondary, rather than their total mass, we use the individual rather than
the system mass function. To ensure that not using the system mass
relationship would not significantly affect our numerical results, we
used it in conjunction with our Monte Carlo modeling, finding that its
numerical results were within the errors of those from the standard
Chabrier distribution. We normalize all three of these mass
distributions according to the mass range they cover, from 0.5 to
1.0\,\msun.

\begin{figure}
\centering \includegraphics[width=\textwidth]{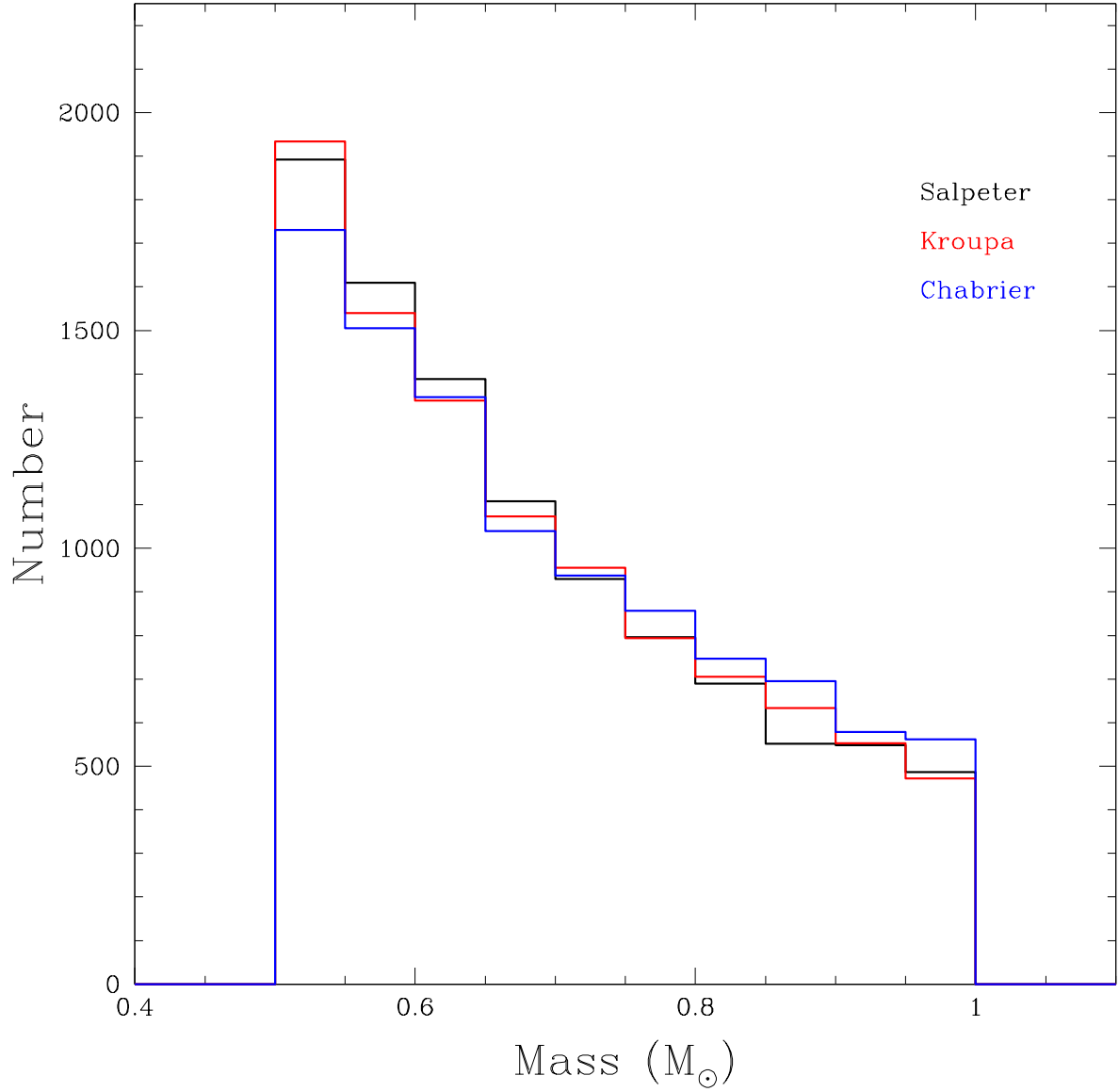}
\figcaption[primary_dist.eps]{The three mass distributions for the
primary stars. They are quite similar to one another, resulting in the
primary mass distribution being largely model independent.
\label{fig:primary_dist}}
\end{figure}

Despite differences in form, these three distributions are, in actuality,
quite similar to one another (see Fig.\,\ref{fig:primary_dist}), making
it largely irrelevant which of the three we choose. When not explicitly
stated, we use the Chabrier distribution, because it was defined
specifically for our mass range of interest.

\subsubsection{Secondary Mass Distribution} 
The mass distribution for the secondary depends on both the mass of the
primary and the period of the system. We use primary-constrained
pairing, i.e. the mass of the secondary is determined from a specified
distribution and is limited by the mass of the primary \citep{kouwen08};
more specifically, the mass of the secondary is between 0.1\,\msun\, and
M$_{prim}$.  The photometric uncertainties of SDSS are 2\% for $gri$ and
3\% for $u$ and $z$ \citep{ivezic04, abazajian05}. For our binaries, the
photometric changes from blending with a secondary are significant,
i.e. greater than the uncertainties, for a primary of 0.5\,\msun\, when
it is blended with a 0.2\,\msun\, secondary for a change in $g$ and $r$
of $\approx$0.05$-$0.1, depending upon the metallicity.  By expanding
our model to secondaries with masses of 0.1\,\msun\,, we ensure that we
cover the entire mass range where a companion can influence the
parameters, both photometric and spectroscopic, of its primary.

Work by \citet{abtwill92} on $\sim$70 binaries with F-G type primaries
determined that for those with short period, defined as less than 100
years, the distribution of secondary masses is flat, confirming previous
work using a sample of 94 binaries with solar-type primaries by
\citet{abtlevy76}. This is of particular interest, because
\citet{hurley05} find that having a flat mass distribution for short
period binaries results in more blue stragglers in their models of M67,
helping them better match observations. \citet{dm91}, however, hesitate
to adopt a flat mass distribution for short-period binaries, while
acknowledging the ambiguities resulting from their small sample size of
$\sim$80 pairs. Currently, there is not firm observational evidence for
or against a flat secondary mass distribution for pairs with short-periods, as the
sample sizes analyzed have been too small to determine the distribution
conclusively, and future work must be done on much larger observed samples.

\begin{figure} 
\centering \includegraphics[width=\textwidth]{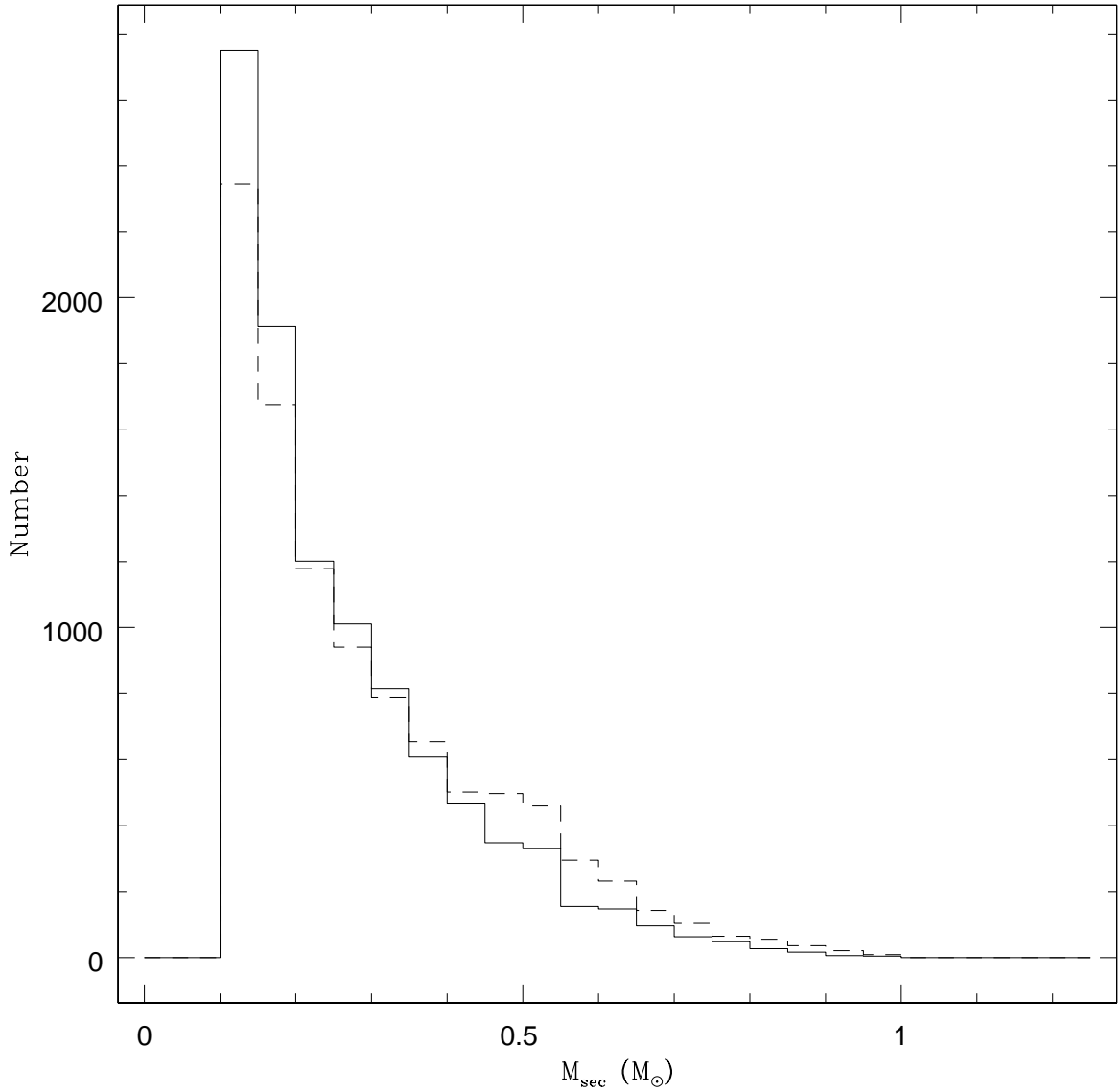}
\figcaption[flat_vs_reg_massdist3.eps]{The Chabrier mass
distribution of secondaries (solid line) compared to the same mass
distribution when all pairs with periods less than 1000 days follow a
flat mass distribution for secondaries (dashed line).  The short
period effect flattens the secondary mass distribution, for all of the
different models.
\label{fig:flattens}}
\end{figure}

For our models of short periods, defined as less than 1000 days based on
the criteria of \citet{dm91}, we try both a flat distribution, as
determined by \citet{abtwill92}, and one that follows the behavior
defined for long periods. We model this for all of our secondary mass
distributions. Defining a flat mass distribution for short-period
binaries affects approximately 22$\pm$1\% of all G-K dwarf blends by
moderating the secondary mass distributions. For all of the
distributions, the number of low mass secondaries are decreased and the
number of high mass are increased. Despite this, the basic overall shape
of the distribution is not affected (see Fig.\,\ref{fig:flattens}).  It
also does not affect the numerical results. The number of blended pairs,
etc.  are within the uncertainties of one another for simulations with
and without the short-period modifications.  For the remainder of this
analysis, we assume that short- and long-period binaries have the same
secondary mass distribution, independent of the system's period.

We employ a variety of distributions to model the secondary masses,
including those used to model the primary masses, the Salpeter, Kroupa,
and Chabrier distributions. All secondary mass determinations are
normalized assuming a mass range from 0.1\,\msun\, to the mass of the
primary. Note that the Kroupa distribution has a different exponent at
masses less than 0.5\,\msun\,:
\begin{equation}
f(M)\propto M^{-1.3} \textrm{ if }0.08 \leq M \leq 0.5
\end{equation}

Additionally, we adopt two relationships based upon \emph{q}, the mass
ratio between the secondary and primary. The first is a Gaussian
distribution from \citet{dm91}:
\begin{equation}
f(q)\propto \exp[-(q-\mu)^{2}/2\sigma^{2}]
\end{equation}
where $\mu= $0.23 and $\sigma= $0.42. The second \emph{q} ratio
distribution we use is from \citet{halb03}. This model takes into
account the prevalence of ``twins," binaries with two stars of
approximately equal mass. Similarly to \citet{dm91}, it is derived
empirically from CORAVEL data, but covers a slightly wider range of
spectral type, from F7 to K, and includes cluster stars in addition to
targets in the solar neighborhood. This particular model has three
peaks, at \emph{q}=0.25, 0.65 and 1.  

\begin{figure}
\centering \includegraphics[width=\textwidth]{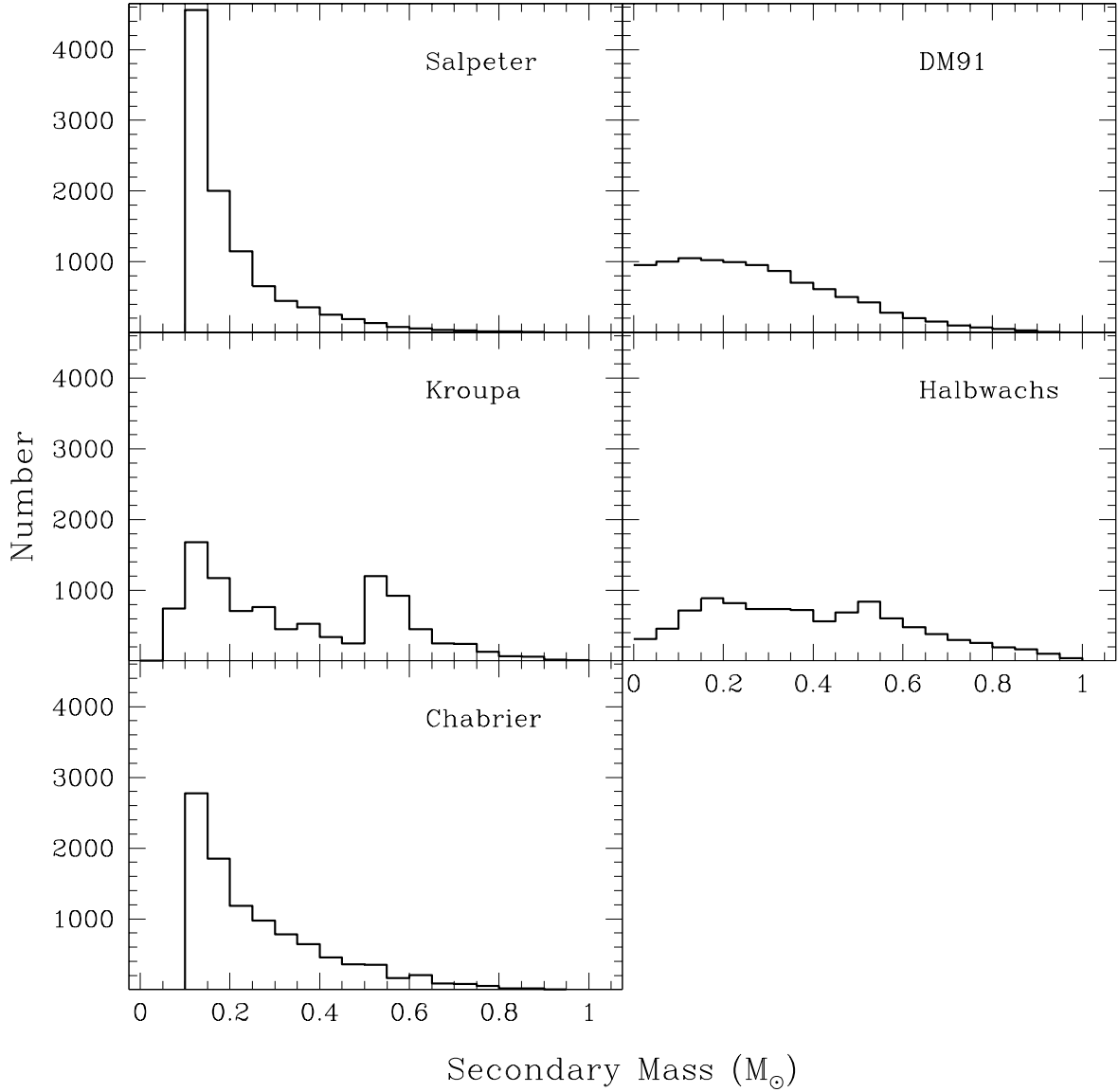}
\figcaption[secondary_dist.eps]{The five mass distributions for the
secondaries. Their differences result in the variation of numbers of
photometrically and spectroscopically blended binaries.
\label{fig:secondary_dist}}
\end{figure}

These five different models have very different distributions, as
expected and shown in Fig.\,\ref{fig:secondary_dist}. By using all of
these different methods, we can compare the results and measure the
effect of using a mass function rather than an empirical \emph{q}
distribution to define the secondary mass distribution. We discuss this
further in \S\,\ref{sec:numbcomp}.

\subsubsection{Eccentricity}

We adopt an eccentricity distribution from \citet{dm91}. If the period
is less than the circularization period, $\approx$10 days, the orbit is
circularized by tidal interactions and has an eccentricity, \emph{e}, of
0.  The eccentricity of tight binaries, with 10$\leq$P$\leq$1000 days,
has a Gaussian distribution, with a mean of 0.31 and a dispersion of
0.155. Again, these values were determined from a sample of nearby G
dwarfs; similar mean values have been determined for young open clusters
and halo stars as well \citep{dm91} .  Lastly, we consider the eccentricity of
long-period binaries, with P$>$1000 days.  \citet{dm91} determined that
for these binaries,
\begin{equation}
f(e) \propto 2e
\end{equation}
Using these three distributions, we assign an eccentricity to each
sample pair based upon its period.

\subsubsection{Inclination}
We assume a flat distribution of $\cos I$ between 0 and 1, where
\emph{I} is the inclination angle. This assumes that all orientations of
the binary system in space are equally likely.

\subsubsection{Phase}

The position of the secondary with respect to the primary determines
whether or not the two will be blended within the SDSS PSF.  For each
secondary we pick a random mean anomaly, \emph{M}, between $-\pi$ and
$\pi$. Combining this with the assigned eccentricity, we calculate the
eccentric anomaly, \emph{EA}, using iterations of a Taylor series
expansion of the Kepler equation \citep{md00}.
\begin{equation}
f(EA)=EA-e\sin(EA)-M
\end{equation}
This results in each pair having an \emph{EA} between 0 and 2$\pi$.  For
pairs with non-negligible eccentricities, the secondary will spend a
larger fraction of its time in orbit far away from the
primary. Determining the parameter \emph{EA} from \emph{M} takes this
distribution into account. From the eccentric anomaly, we then calculate
the true anomaly, \emph{F}.
\begin{equation}
\tan(F/2)=\sqrt{\frac{1+e}{1-e}}\tan(EA/2)
\end{equation}
With \emph{F}, we determine \emph{r}, the distance between the two stars
at that particular orbital phase.
\begin{equation}
r=\frac{a(1-e^2)}{(1+e\cos(F))}
\end{equation}
where \emph{a} is the semimajor axis of the orbit, which we derive from
the period and masses of the pair. Finally, we select random values
between 0 and 2$\pi$ for $\Omega$, the longitude of the ascending node,
and $\omega$, the argument of pericenter.  We use \emph{r} in
conjunction with the true anomaly, \emph{F}, $\Omega$, and $\omega$ to
determine the two orthogonal components of the projected distance
between the secondary and the primary, $X$ and $Y$.
\begin{eqnarray}
X &=& r[\cos\Omega \cos(\omega+F) - \sin\Omega \sin(\omega+F) \cos I] \\
Y &=& r[\sin\Omega \cos(\omega+F) + \cos\Omega \sin(\omega+F) \cos I]
\end{eqnarray}
Using $X$ and $Y$, we then calculate the projected
magnitude of the separation on the sky of the two stars, R:
\begin{equation}
R = \sqrt{X^2 + Y^2}
\end{equation} 

\subsection{Photometric Blending}
\label{sec:distance} 
As mentioned earlier, for a pair to be blended photometrically, they
must be within 1.4\arcsec\, of each other on the sky. Combining the
orbit information with an assigned distance, we can determine whether or
not the pair fulfill the separation criteria. If blended, we then
combine $ugriz$ magnitudes, based on isochrones.

To model the distances to each pair, we determine an empirical distance
distribution for G and K dwarfs with spectra from SEGUE using their
target selection parameters in the $(g-r)_0$ color and $r_0$ magnitude
\citep{dr6}. From this sample, we first eliminate all targets for which
the SSPP was unable to determine a metallicity, temperature, or
$\log\,g$.  In total we extract approximately 20,000 G-K stars from
SEGUE, two-thirds of which are SEGUE-defined G type.

\begin{figure}
\centering \includegraphics[width=\textwidth]{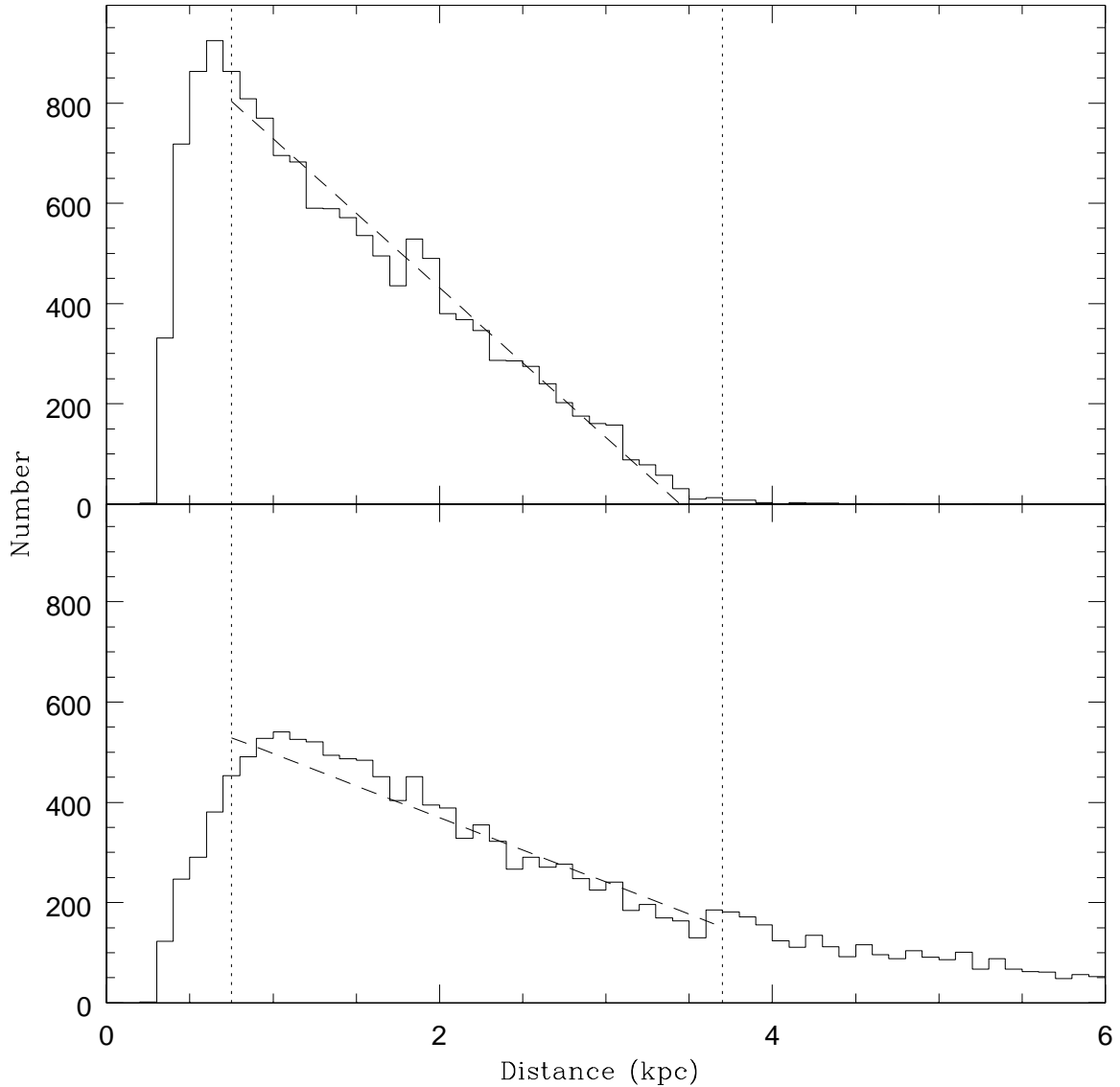}
\figcaption[distance_distributions2.eps]{The distance distribution of the 
SEGUE G-K dwarf sample (top). Each target is compared to the Dartmouth isochrones 
based upon its SSPP temperature determination. We then use the observed \emph{g} and 
\emph{r} magnitudes in conjunction with the isochrone to determine a distance to 
each target. To ensure that we cover the same volume range for both G and K dwarfs, 
we limit the distance range to be between $\sim$750 and $\sim$3700 pc, as indicated 
by the short dashed lines. The long dashed line is a least squares linear fit to the 
histogram of distances (see equation\,\ref{eq:distorig}). 
The lower plot is the spread of distances once we have contaminated the original sample 
with binaries. A randomly selected 65\% of the SEGUE targets, the expected binary frequency according 
to \citet{dm91}, are given distances that are twice as far as those originally determined by comparison 
with the isochrones. This significantly changes the slope of the least squares fit (see 
equation\,\ref{eq:distcont}). We use both of these distance relationships in our Monte-Carlo fit, and 
find that the difference in distance distributions resulting from binary contamination in the SEGUE sample
has a negligible effect on the numerical results, i.e. the number of photometric and spectroscopic blends. 
\label{fig:distdist}}
\end{figure}

For each SDSS target, we then select a 10 Gyr comparison isochrone from the
Dartmouth set based upon the SSPP metallicity over a range of
[Fe/H]=$-$0.5 to $-$2.5 (see \S\,\ref{sec:isochrones} for more
information on the isochrone selection).  If the target's metallicity
falls between two isochrones, we interpolate.  We then match the target
to the isochrone by SSPP temperature and pull out the modeled
\emph{ugriz} magnitudes, surface gravity, and mass from the isochrone.
Comparing the isochrone magnitudes to those detected by SDSS, we use the
distance modulus to determine an approximate distance to each target
based on the $g_0$ and $r_0$ magnitudes, weighted equally. For this
sample we find a range of distances from 0.5 to 6 kpc (see
Fig.\,\ref{fig:distdist}). As G dwarfs are brighter than K dwarfs, SDSS
can observe them at greater distances. Based on the magnitude limits of
SDSS, we select a distance range from $\sim$750 and $\sim$3700 pc.  to
ensure that our G and K dwarf sample occupy the same volume of space.
We fit the distribution of distances in this range using a linear least
squares fit, measuring:
\begin{equation}
f(d) = -298(d/kpc)+1027
\label{eq:distorig}
\end{equation}
Each modeled pair is thus assigned a distance according to this
distribution, and, in conjunction with each pair's R value, we derive a
projected separation on the sky in arcseconds.

If our sample of dwarfs from SEGUE is contaminated by binaries, the
distance distribution will be affected; targets will appear brighter,
and we will underestimate their distance. To simulate this effect, we
adjust our measured distances for the G-K dwarfs in SEGUE. We select a
random 65\%\, of our sample, the percent expected to be in binaries
\citep{dm91}, and double their calculated distance. As undetected
secondaries will cover a range of masses, by assuming that all of the
pairs are twins, we calculate the most extreme scenario of binary
distance contamination. We again determine a linear least squares fit to
the distribution of distances, which now has a much shallower slope:
\begin{equation}
f(d) = -128(d/kpc)+625 
\label{eq:distcont}
\end{equation}
For the G-K color pairs, using the most extreme contaminated distance
relationship results in an additional $\sim$2 percentage points of
blends, independent of metallicity and mass distributions. Consequently,
we feel confident using the original distance relationship and ignoring
the potential binary contamination effects.

\subsection{Spectroscopic Blending}
We have determined the combined magnitudes of pairs photometrically
blended in SDSS photometry. However, as the spectroscopic fiber for
SEGUE is 3\arcsec, we must take into account that there are some pairs
that, although not photometrically blended, are spectroscopically
blended.

Spectroscopic blends are photometrically distinguishable. If SEGUE
recognizes them as close pairs and systematically avoids them, due to
concerns about spectral contamination, we do not need to take
spectroscopic blends into account.  The SDSS photometry pipeline has a
deblending function capable of resolving a binary with a separation of
1\arcsec\, or greater (Yanny, private communication). If the deblend is
not done well, as indicated by certain photometric flags, SEGUE avoids
that particular target. However, a well-resolved pair will not be
avoided by SEGUE. As SDSS can separate stars as close as $\sim$1\arcsec,
this indicates that SEGUE will likely include many binaries with
separation less than 3\arcsec\, in the sample, and the primaries will
have their spectroscopic parameters contaminated by a secondary. Thus,
we count every pair with 1.4\arcsec$\leq r \leq$3\arcsec\, as a
spectroscopic blend capable of contaminating the SEGUE sample and
affecting SSPP measurements.

\section{Number Comparison}
\label{sec:numbcomp}

We use a Monte Carlo method to determine the effect of binaries on the
G-K spectroscopic sample in SEGUE. The first step is to determine how
frequently a photometric blend will cause a binary that would otherwise
fall into the G-K sample to fall outside the color range, or how often a
photometric blend will put a pair into the sample whose primary would
otherwise be too blue. We examine whether the fraction that shift into
or out of the color range depends on the primary or secondary mass
function.  For each combination of metallicity and mass distributions,
we pick 100,000 primaries drawn from the primary mass distribution and
match them with 100,000 secondaries, drawn from the secondary mass
distribution. These pairs are given orbits in accordance with the
previous discussion (see \S\,\ref{sec:binary properties}).  As shown in
Fig.\,\ref{fig:primary_dist}, the three different mass distributions for
the primaries result in approximately the same number of stars at each
mass. However, the mass distributions for the secondaries have very
different forms (see Fig.\,\ref{fig:secondary_dist}).  For each combination
of primary and secondary mass distribution, we determine the number of
pairs that are photometrically (projected closer than 1.4\arcsec) and
spectroscopically (projected closer than 3\arcsec) blended.

\begin{figure}
\centering \includegraphics[width=\textwidth]{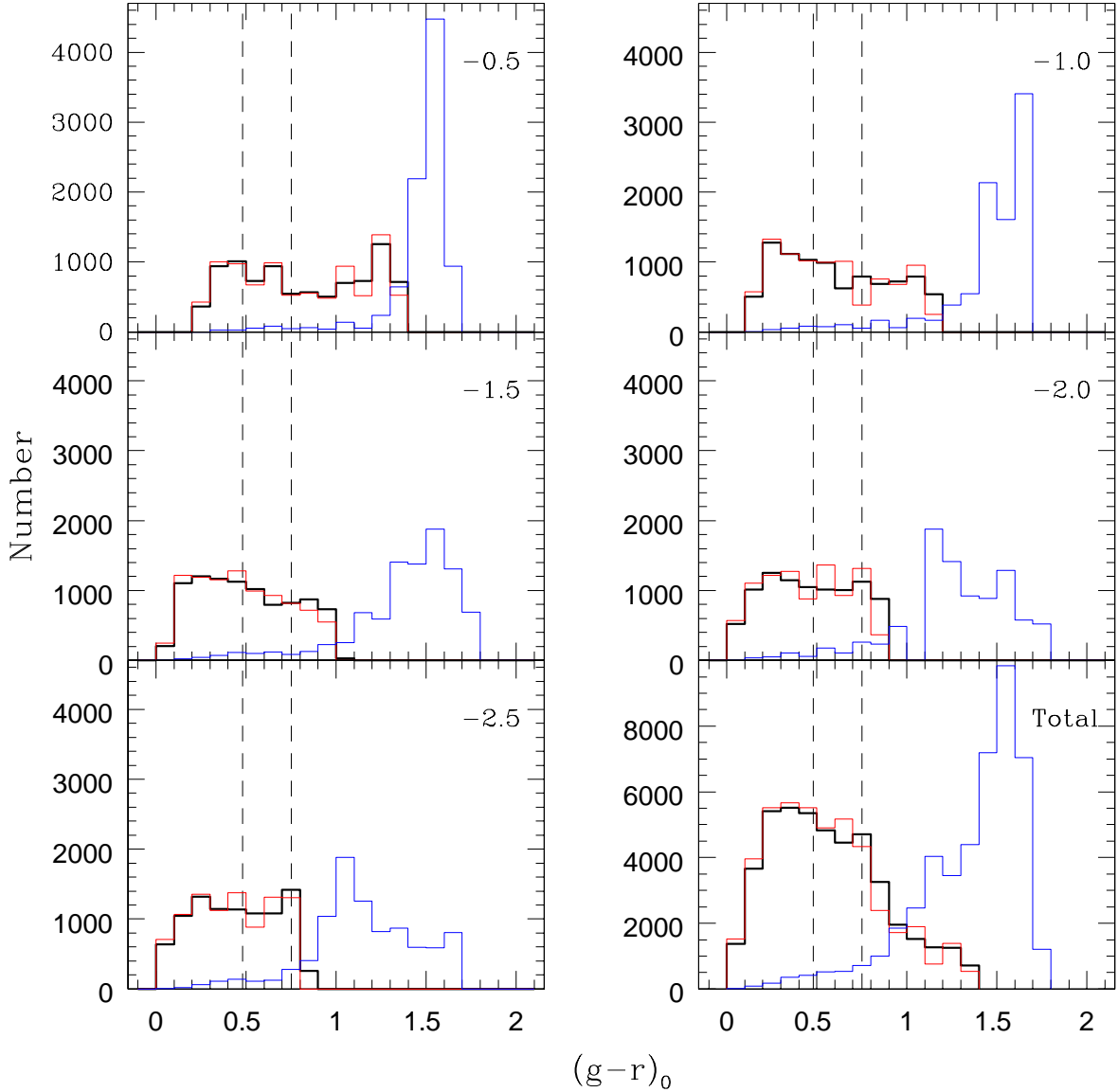}
\figcaption[cc_gminr_update2.eps]{Histograms of the shift in $(g-r)_{0}$ color over
the entire blended sample. The red histogram represents the primaries,
the blue is the secondaries and the black is the binaries. The dashed
lines indicate the G-K dwarf color cut as specified by SEGUE target
selection \citep{yanny09}.  At the top right corner of each plot is the
metallicity for that sample; the distributions shift slightly with
metallicity. The combined sample is in the bottom right plot. Note that
the numbers of targets in the color cut range does not vary
significantly from the primaries to the pairs, indicating that the
addition of a secondary will not have a large affect on the population
numbers in the G-K range.  This model assumes a Chabrier primary and
secondary mass distribution, although the behavior is similar for the
range of mass distributions. \label{fig:gminr} }
\end{figure}

Of a sample of 100,000 pairs, approximately 90$\pm$1\% are
photometrically blended, regardless of mass distribution or metallicity.
Approximately 40\% of all targets will have their $(g-r)_0$ color
shifted by an amount greater than the uncertainties in SDSS photometry
by the addition of a secondary.  We then use color cuts to extract the
pairs that will be within the $(g-r)_0$ color range of G and K
dwarfs. These are the binaries that will be contaminating the G-K dwarf
sample and possibly affecting SEGUE target selection, resulting in
potential errors in studies of stellar populations. Similar to the
larger sample, $\approx$40\% of the G-K dwarf sample will have their
$(g-r)_0$ color shifted by more than the SDSS photometric
uncertainty. In Fig.\,\ref{fig:gminr}, we compare the $(g-r)_0$ color
distribution for the primaries to that of the blended pairs over a range
of metallicities, examining the numbers of targets that remain within
the G-K dwarf color cut.  On average, 30$\pm$2\% of the 100,000
primaries are within the G-K color cut, compared to 28$\pm$2\% of
the pairs (see Table\,\ref{tab:numbers}). The addition of a
secondary bumps out slightly more targets from the G-K range than it
bumps in.  On average, 2\% of the 100,000 primaries will be
bumped into the color cut by adding in a secondary; 3\% will be
pushed out of the $(g-r)_0$ range by a companion.  In addition to
calculating percentages, we examine the most frequent color shifts by
comparing the $(g-r)_0$ of the blended binaries to those of the
primaries. We analyze a sample which includes every combination of
primary and secondary mass distribution at every metallicity. For the
entire sample, the addition of a secondary most often shifts $(g-r)_0$
by 0.01 with $\sigma$=0.02. When we extract all of the targets for which
the primary is in the G-K dwarf SEGUE color range, the shift and
$\sigma$ are the same. The population error resulting from
photometric binary contamination in target selection is minimal, as the 
shifts themselves are small. 

We isolate G-K dwarf spectroscopic blends by selecting pairs where the
primary is within the color cuts. As noted earlier (see
\S\,\ref{sec:pair_modeling}), using 3\arcsec\, as the criteria is the
most extreme scenario for spectroscopic contamination. On average, about
3\% of the G-K sample are spectroscopically, but not
photometrically, blended. Thus, even for the largest possible chance of
purely spectroscopic contamination, the likelihood is negligible.

\begin{figure}
\centering \includegraphics[width=\textwidth]{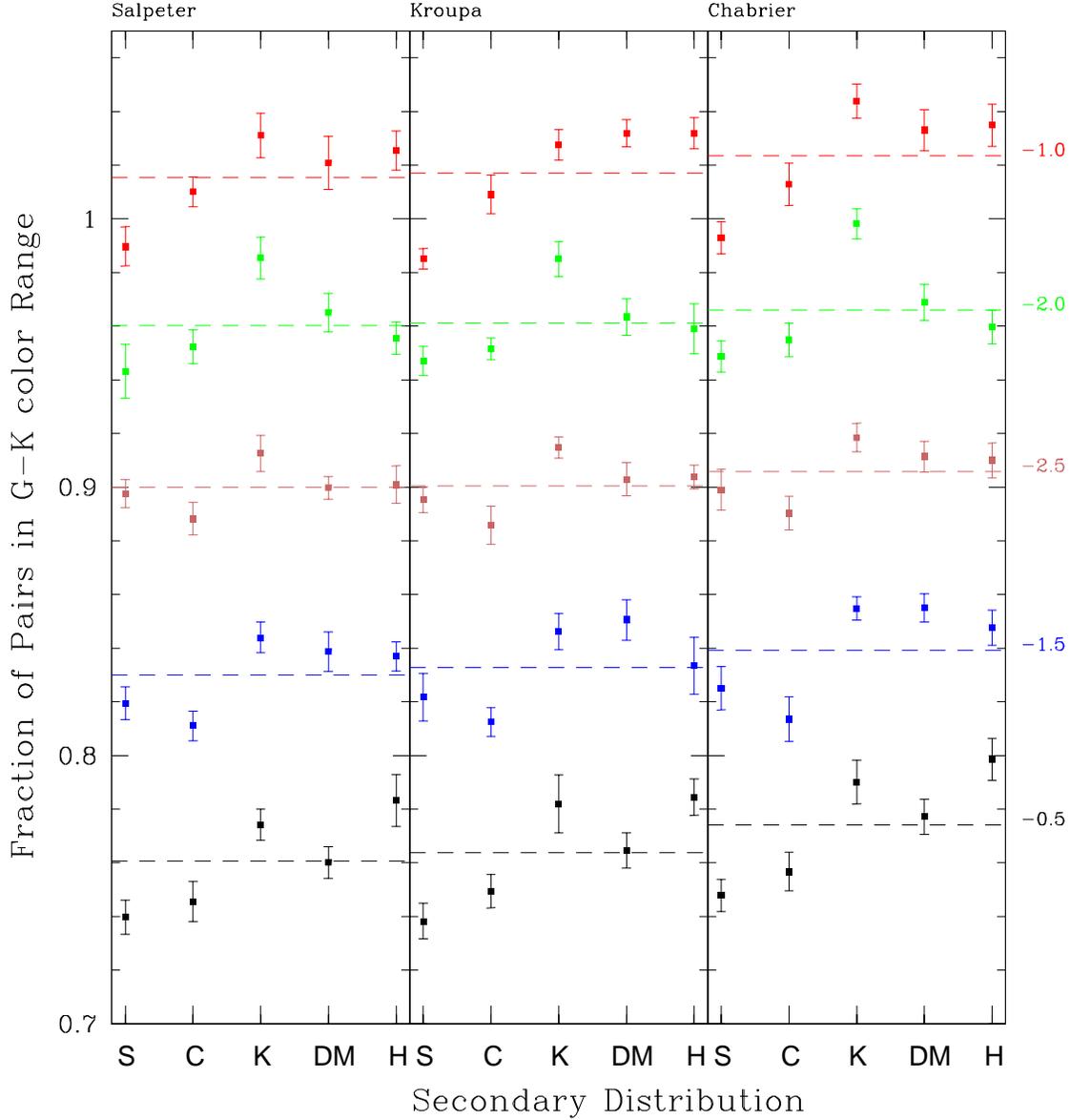}
\figcaption[pairs_sec_frac_errors.eps]{The fraction of pairs that are
blended spectroscopically and photometrically in the sample of G-K range
binaries.  The three panels represent the three different mass
distributions of the primaries, Salpeter, Kroupa, and Chabrier. The x
axis is labeled with the various mass distributions of the secondaries:
Salpeter (S), Chabrier (C), Kroupa (K), Duquennoy \& Mayor $q$ ratio (DM),
and Halbwachs $q$ ratio (H). Each color represents a different
metallicity.  Black is [Fe/H]=$-$0.5, red is $-$1.0, blue is $-$1.5,
green is $-$2.0, and pink is $-$2.5.  Each combination of primary and
secondary mass distributions was run 10 times; the error bars reflect
the RMS variation in the blended fraction for the total 100,000 target 
sample.  The average fraction for all
of the secondary distributions is noted with the dashed line.  As
expected, the fractions for the Salpeter and Kroupa primary mass
distributions agree, while the Chabrier is slightly
different. Additionally, all but a few of the fractions for each of the
secondary distributions agree with one another within the RMS
uncertainties, indicating that the fraction of blended binaries is
approximately independent of the secondary mass distribution.
\label{fig:pairs_sec}}
\end{figure}

In Fig.\,\ref{fig:pairs_sec}, we plot the fraction of blended pairs
within the G-K dwarf color range, both photometric and spectroscopic,
versus the secondary mass distributions for all three primary
distributions over the range of metallicities. As listed in
Table\,\ref{tab:numbers}, we examine how many primaries are initially in
the G-K color range and compare it to the number of pairs within the
color cut.  We plot the average blend number with the error bars
representing the standard deviation among the 10 samples of 10,000. As
expected, there is variation in the blended fraction related to the
different secondary distributions. However, in general, the statistics
for the range of secondary distributions agree with one another within
uncertainties for each primary mass distribution, indicating that there
is not a substantial numerical difference from one model to another (see
Table\,\ref{tab:numbers}).  We are not surprised by this because
different factors, such as the short period flattening and Chabrier
system mass distribution, appeared to have little to no effect on the
number of resulting blends as well. The fractions determined for the
Salpeter and Kroupa primary distributions are quite similar to one
another, whereas the Chabrier results are slightly increased.

Recent work by \citet{met09} determined that the mass functions of
companions are different than those of individual objects, making it
inaccurate to use the empirical mass functions, such as the Salpeter,
Kroupa, and Chabrier relationships, to model and analyze stellar
secondaries. 
However, \citet{met09} find good agreement between the distribution of
companion masses and the relationship measured by \citet{dm91} for all
targets with \emph{q}$>$0.1. We do not find significant variation in our
blended G-K fraction with different secondary models. The model using
the \emph{q} relationship from \citet{dm91} appears quite similar
numerically to the other models, where we pull random secondary masses
from an IMF.  Thus, we are not particularly concerned about the detected
differences between the primary and secondary mass function, as it does
not appear to have a significant effect on our statistical modeling.

\begin{figure}
\centering \includegraphics[width=\textwidth]{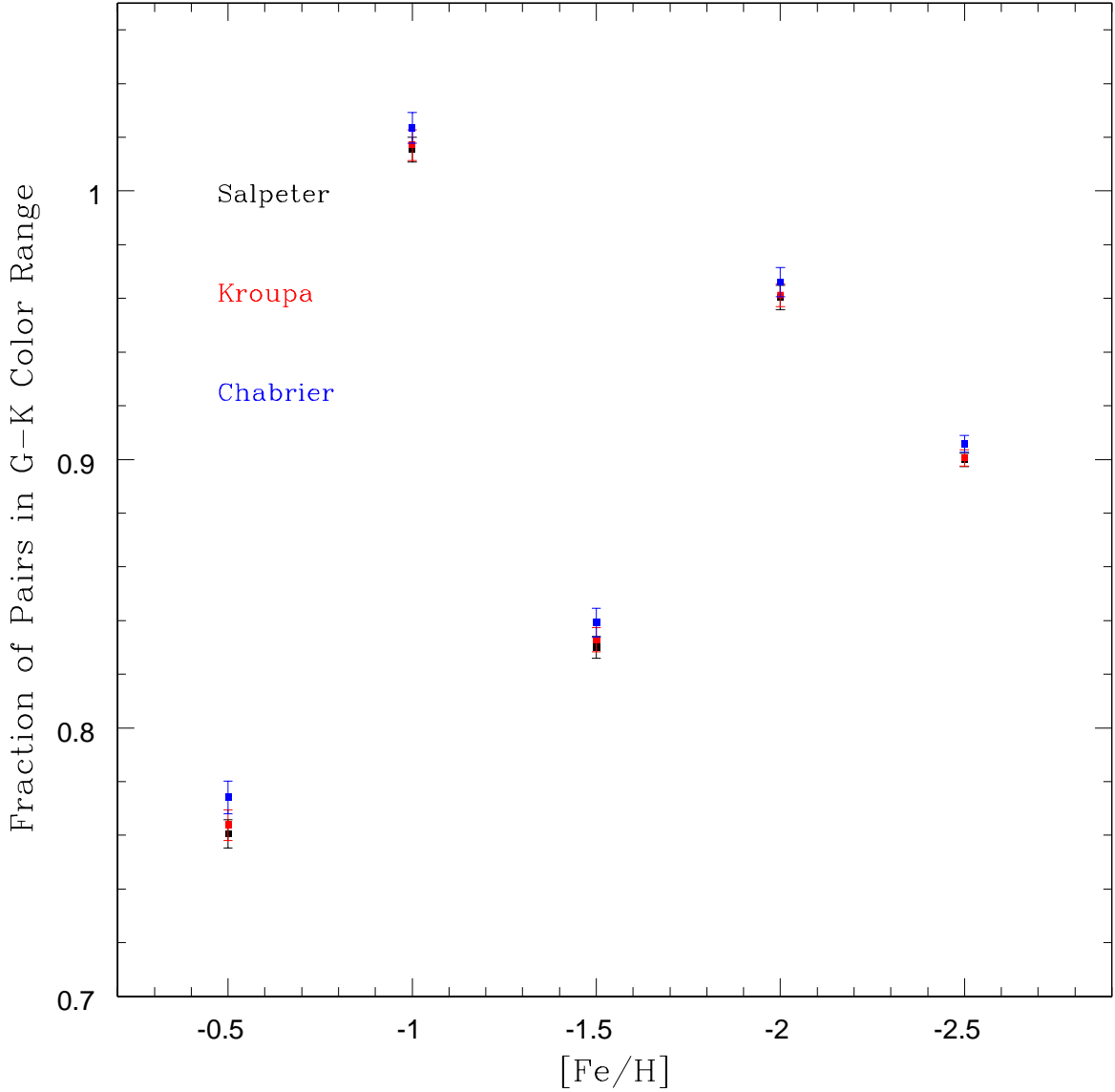}
\figcaption[pairs_met_frac_errors2.eps]{The average fraction of blended pairs
for each metallicity. The black points have a Salpeter primary mass
distribution, the red have Kroupa, and the blue represent
Chabrier. There is no particular trend of fraction of blends with
metallicity; the differences are the result of our modeling scheme (see
\S\,\ref{sec:numbcomp}).  Note that the fraction of blends agrees
quite well for Salpeter and Kroupa but is slightly different for
Chabrier, which is slightly higher, as we saw in
Fig.\,\ref{fig:pairs_sec}.  The uncertainties are the RMS variation for
the average fractions designated by the dashed lines in
Fig.\,\ref{fig:pairs_sec} determined from the complete sample. \label{fig:pairs_met}}
\end{figure}

For each primary distribution, we plot the average fraction of blends at
each metallicity and compare them in Fig.\,\ref{fig:pairs_met}.  There
is no clear relationship between the blended fraction and metallicity.
The fractional variation over our metallicity range is likely due to our
color cut, rather than a real physical effect. At each metallicity, the
mass range of our pairs covers a different $(g-r)_0$ color range. Lower
metallicity primaries cover a smaller range of colors which are shifted
more towards the blue. Conversely, the spread in $(g-r)_0$ color range
of the secondaries is larger for lower metallicities. The different
spread of colors in both primaries and secondaries changes the effect a
secondary has on the primary's color, in particular whether or not it
can shift primaries in and out of the G-K color range.  This causes
variation, but not a consistent trend, in the fraction of blends that
remain within the G-K color range with metallicity. Independent of the
effects of metallicity, all of the blended fractions determined by the
different distributions agree with one another within the errors,
indicating that our choice of mass distributions has little effect on
our numerical results.

\section{Spectroscopic Modeling}

To understand the effect of simulated binary contaminants on SSPP
parameter determinations, we spectroscopically model a grid of binaries
over the range of metallicities and process them through the
pipeline. Each member is of a particular mass, with the temperature,
surface gravity, luminosity, and \emph{ugriz} magnitudes extracted from
the Dartmouth isochrones. We then model the spectra using a MARCS model
atmosphere grid \citep{gustaffson08} in conjunction with the
TurboSpectrum program using these Dartmouth parameters \citep{ap98}.

\subsection{Model Atmosphere Grid}

We utilize MARCS model atmospheres of standard chemical composition and
plane parallel geometry to develop a grid of model spectra, covering a
range of temperature, surface gravity, and metallicity
\citep{gustaffson08}. Each spectrum in the grid is interpolated from
MARCS model atmospheres using the bracketing values in effective
temperature, surface gravity, and metallicity. For our interpolation,
the temperature is in increments of 25 K from 3200 to 8000, $\log g$ in
0.1 dex from 3.0 to 5.5, and metallicity in steps of 0.5 ranging from
[Fe/H]=$-0.5$ to $-2.5$.  For standard composition models of
[Fe/H]=$-0.5$, [$\alpha$/Fe] is defined as 0.20.  For all other
metallicities, [$\alpha$/Fe]=0.40. Finally, our microturbulence is
defined as 2 km/s.

\subsection{Model Synthesis}

\begin{figure}
\centering \includegraphics[width=\textwidth]{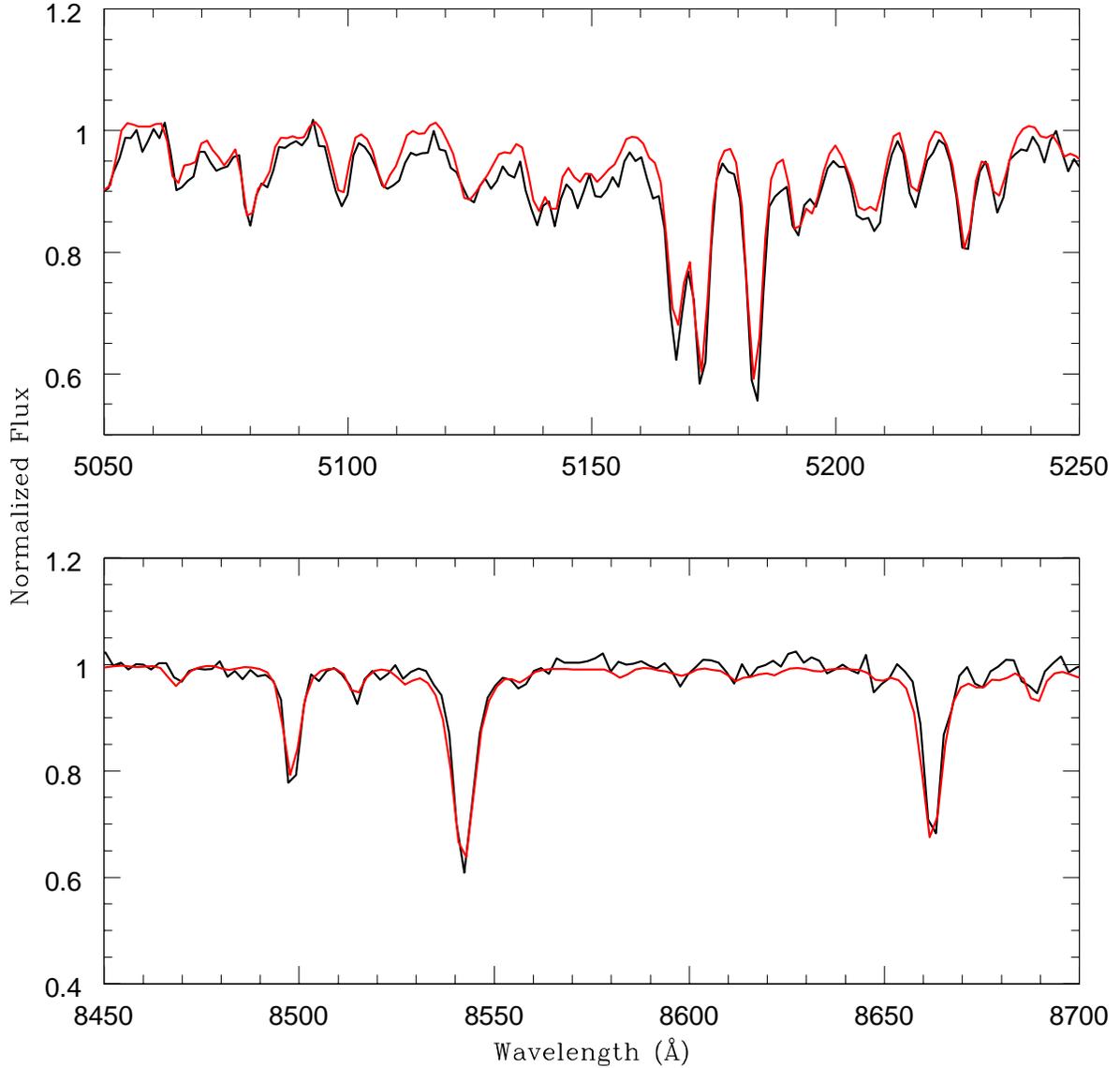}
\figcaption[speccomp_ed3.eps]{A processed model spectrum (red) vs. an actual
SEGUE spectra (black).  We have focused in on two regions for
comparison: the region of MgH features (top) and the Ca\,II triplet
region (bottom). Both are important for determining parameters in the
SSPP.  The model spectra has been made with TurboSpectrum from a grid of
MARCS model atmospheres. The model is then adjusted to match the
dispersion of SDSS and binned to have 69 km/s per pixel. Note that both of these spectra
have been normalized using SPECTRE continuum division to make them
easier to compare.  The parameters of the model are determined from
Dartmouth isochrones. This particular model has a temperature of 5600 K,
$\log g$ of 4.6, and [Fe/H] equal to $-$0.5. It represents a 0.8\,\msun\,
star.  This SEGUE target has a temperature of 5612 K, a $\log g$ of 4.5,
and [Fe/H] of $-$0.59 according to the SSPP. \label{fig:synth_segue.eps} }
\end{figure}

For each metallicity, we synthesize 10 model stellar spectra,
corresponding to the masses from 0.1 to 1.0\,\msun\, by interpolating a
MARCS model atmosphere and processing it in TurboSpectrum (see
Table\,\ref{tab:parameters}).  These models cover the same wavelength
range as SEGUE, 3800-9200\AA, in 0.1\AA\, increments. To simulate the
spectra of binaries, each primary spectra, from 0.5 to 1.0\,\msun\,, is
combined with all secondaries of lesser or equal mass. TurboSpectrum
provides us with the flux from the star; combining this with the radii
of the stars squared, as determined from the Dartmouth isochrones, we
calculate the luminosity of each star. The ratio of these luminosities
determines how much an undetected secondary will affect its primary. We
then apply an SDSS dispersion file to Gaussian-smooth the synthetic
spectra to model the instrument output.  Lastly, the synthetic spectra
are binned into pixels of width 69 km/s. These modifications ensure both
that our spectra are accurate reflections of data taken by SEGUE, and
that our models can easily be run through the SSPP. We have compared a
synthesized model and a SEGUE spectrum in
Fig.\,\ref{fig:synth_segue.eps}.

\section{Parameter Shifts}

We do not expect our synthetic spectra to be a perfect match to real
spectra, particularly for the strongest features, such as the Balmer
lines, because we do not take into account known difficulties related to
NLTE, 3-D, and chromospheric effects in our models.  Therefore, we are
not surprised that there are differences between the parameters we use
to model the spectra and the parameters that the SSPP derives. The
critical issue is understanding how much simulated binarity changes the
values the SSPP determines for the atmospheric parameters of our sample
of synthetic binaries, independent of modeling errors.

\begin{figure}
\centering \includegraphics[width=\textwidth]{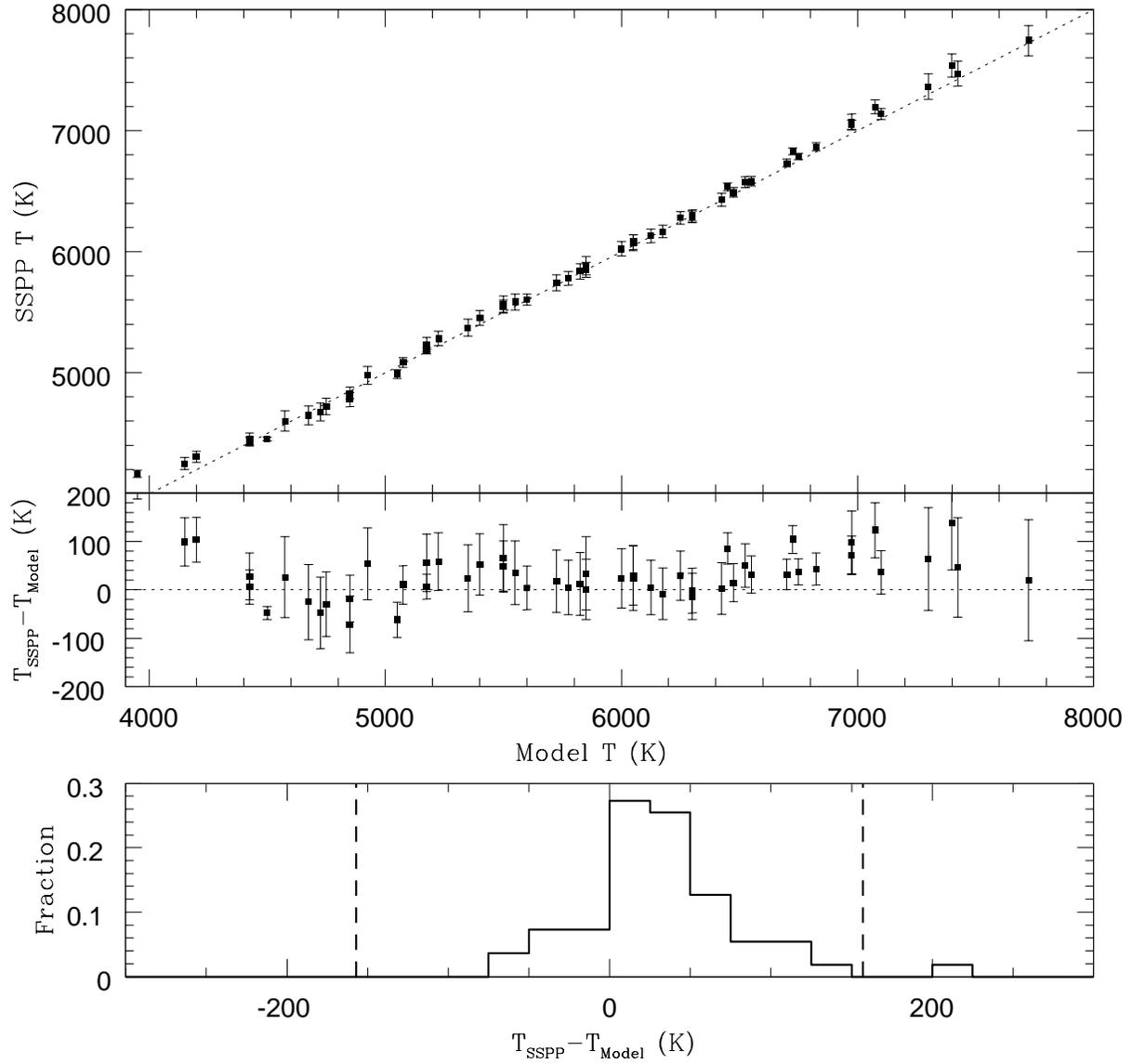}
\figcaption[ssppcomp_prim_temp2.eps]{A comparison of the temperatures of
our primary control sample from 0.5 to 1.0\,\msun\, over a metallicity
range of [Fe/H]=$-$0.5 to $-$2.5. Each of these synthetic spectra was run
through the SSPP. Here we have compared the SSPP temperature output to
the actual temperature set for the model. The SSPP tends to overestimate
the stellar temperatures by $\sim$12 K.
\label{fig:sspp_prim_temp} }
\end{figure}

\begin{figure}
\centering \includegraphics[width=\textwidth]{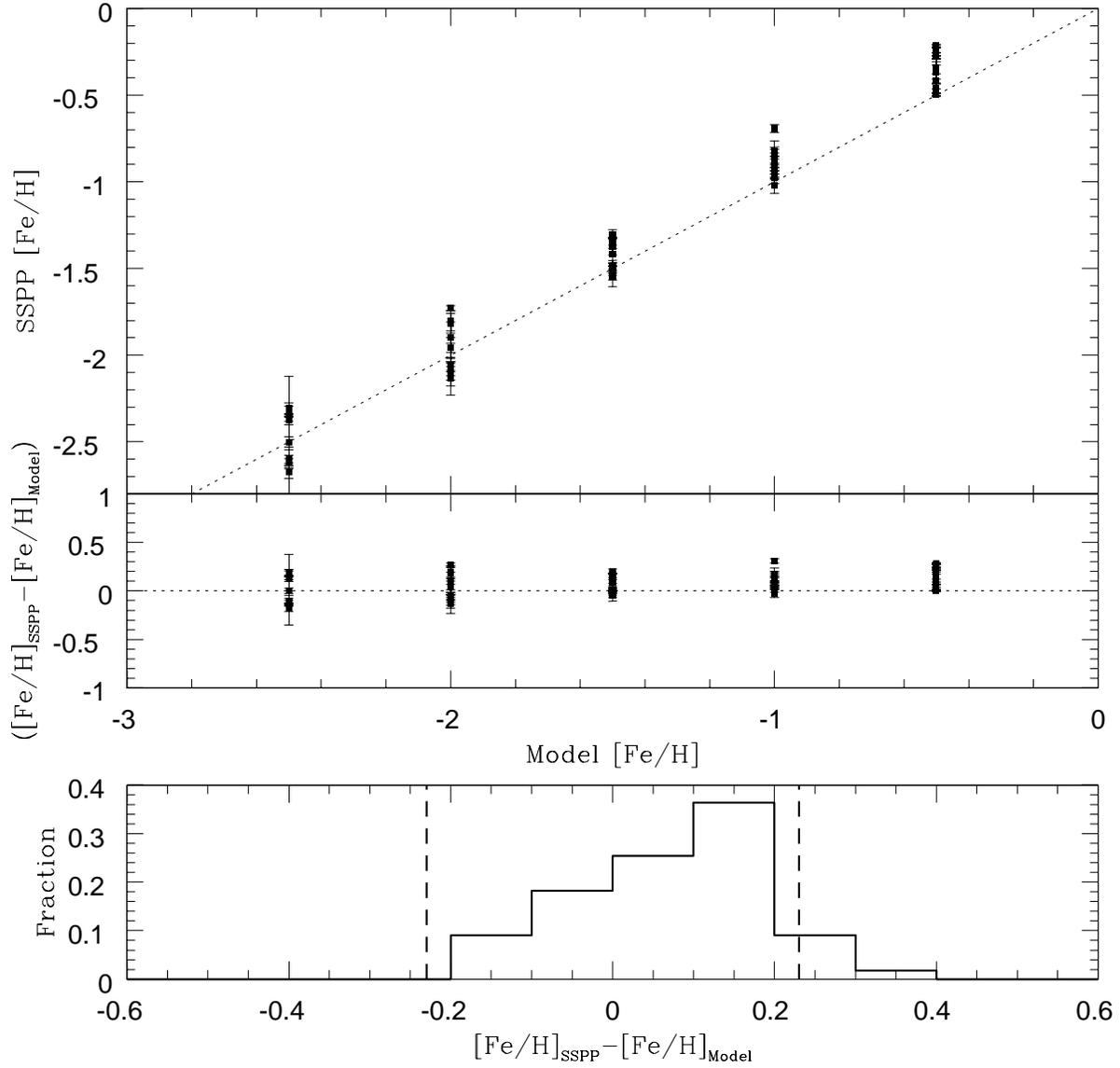}
\figcaption[ssppcomp_prim_feh2.eps]{A comparison of the metallicity of the
synthetic stars control sample. We have compared the [Fe/H] determined
by the SSPP to the values set for the model. The SSPP tends to
overestimate metallicities by $\sim$0.15 dex.
\label{fig:sspp_prim_feh} }
\end{figure}

\begin{figure}
\centering \includegraphics[width=\textwidth]{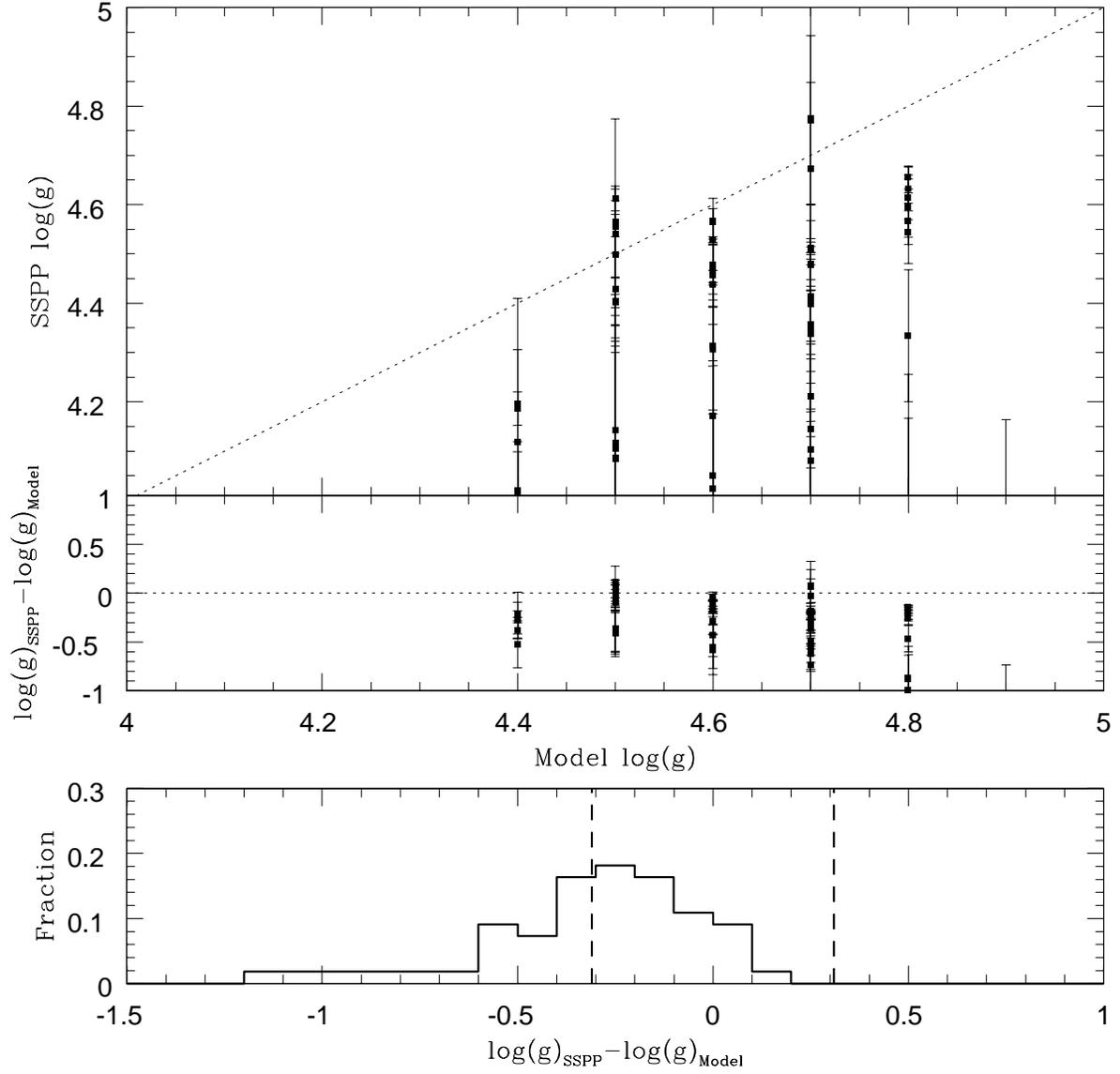}
\figcaption[ssppcomp_prim_logg2.eps]{A comparison of the surface gravity
of the synthetic stars control sample from the SSPP to the actual values
set for the model. The SSPP tends to underestimate the surface gravity
by $\sim$0.25 dex.
\label{fig:ssppcomp_prim_logg} }
\end{figure}

\begin{figure}
\centering \includegraphics[width=\textwidth]{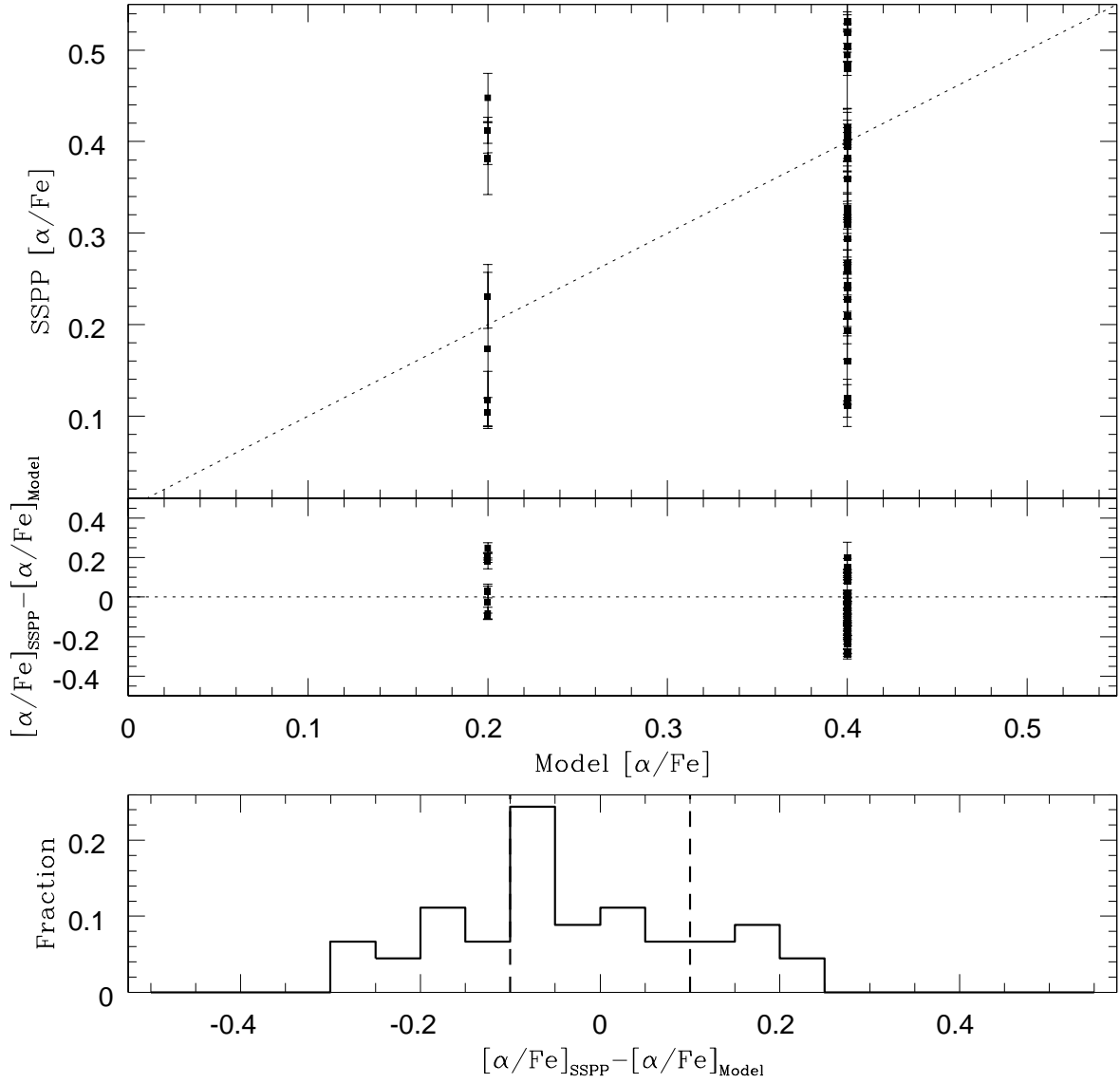}
\figcaption[ssppcomp_prim_alpha2.eps]{A comparison of the [$\alpha$/Fe] of the
synthetic stars control sample from the SSPP to the actual values set
for the model.
\label{fig:ssppcomp_prim_alpha} }
\end{figure}

We run our synthetic spectra of single stars with
0.5\,\msun\,$\leq$M$\leq$1\,\msun\, through the SSPP as a control sample
to determine the SSPP parameter offsets for temperature, metallicity,
surface gravity, and $\alpha$-enhancement. The values measured for the
control group of primaries are listed in
Table\,\ref{tab:control_group}. The temperatures determined by the SSPP
for this sample most frequently overestimate the model temperature by
around 12 K (see Fig.\,\ref{fig:sspp_prim_temp}), a negligible amount
with respect to the expected errors ($\sigma(T_{eff})$$<$150 K) for the
pipeline \citep{lee08_I}. We also compare the [Fe/H] value determined by
the pipeline to that set for the models (see
Fig.\,\ref{fig:sspp_prim_feh}). The pipeline overestimates the
metallicity of the sample with a mode of $\sim$0.15 dex, less than the
expected error of $\sigma([Fe/H])$=0.24 dex \citep{lee08_I}. Whereas
temperature and metallicity are overestimated by the pipeline, surface
gravity is underestimated (see Fig.\,\ref{fig:ssppcomp_prim_logg}). The
most common shift is a decrease of $\sim$0.25 dex, which is comparable
to the expected SSPP error of 0.29 dex. Lastly, we examine the offsets
in [$\alpha$/Fe] in Fig.\,\ref{fig:ssppcomp_prim_alpha}. All of the
synthetic spectra have [$\alpha$/Fe] set to 0.2 or 0.4, based on the
MARCS model atmosphere metallicity. When processed, the SSPP finds a
wide range of $\alpha$-enhancement, spreading the values over a range
from $\sim$$-$0.5 to $\sim$0.25. The most frequent offset is $-$0.075
dex. Since the offsets are smaller than or on the order of the SSPP
errors, we find that our model spectra are more than adequate for
measuring the effects of binary contamination.

Using this control sample of primaries, we define a temperature,
metallicity, and surface gravity offset between the models and the SSPP
parameters due to assumptions in our model synthesis. We then compare
the values determined for the grid of binaries with those determined for
the primary member of each pair. The parameters SSPP determines for each
pair are listed in Table\,\ref{tab:binary_params}, and the differences
between these values and those determined for the primary control group
are in Table\,\ref{tab:binary_diff}. In general, for all of the
parameters, temperature, metallicity, surface gravity, and alpha
enhancement, the shifts between the SSPP determinations for the
primaries and the pairs are most often within the expected SSPP
uncertainties (see Fig.\,\ref{fig:delta_all},
Table\,\ref{tab:binary_diff}). There is a slight anticorrelation between
these values for the [$\alpha$/Fe] with a slope of $-$0.56, but
otherwise the relationships are flat.

\begin{figure}
\centering \includegraphics[width=\textwidth]{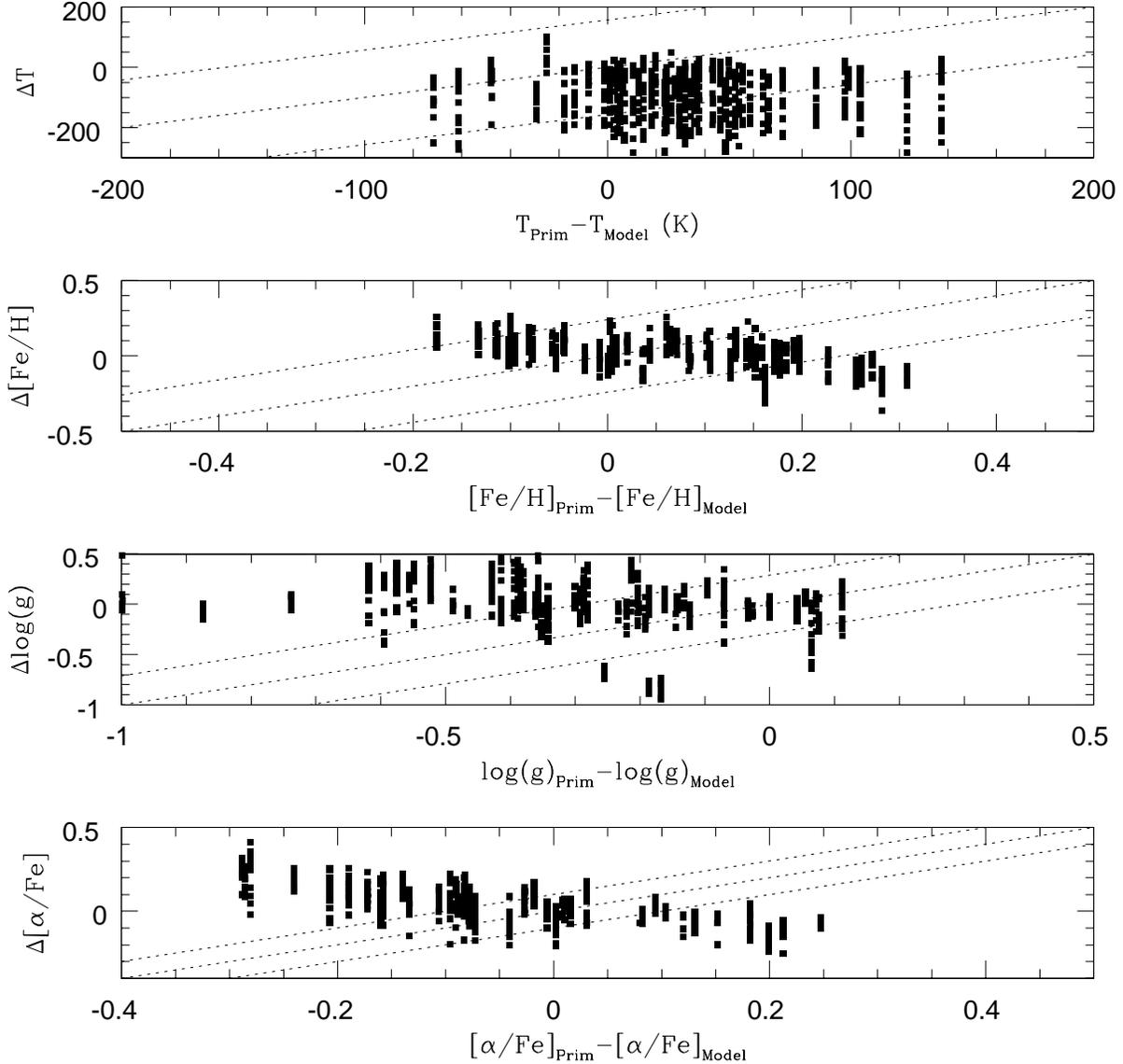}
\figcaption[deltadelta_all3.eps]{Similar to the previous figure, except
the x axis is the difference between the SSPP and models for the
primaries, rather than the SSPP measurements for the control group. The
y axis is the difference between the SSPP determinations for the pair
and that for the primary.  Four parameters from the pipeline are
plotted: temperature, metallicity, surface gravity, and alpha
enhancement. The central dotted line represents a 1:1 correlation
between shifts. Parallel to this line are two dashed lines indicating
the error range for SSPP, i.e. 150 K for temperature measurements.
These diagrams indicate that the shift amounts are not strongly
correlated with each other so the secondary typically has an effect
independent of the standard SSPP offsets, and reinforce that the shifts
are in general small for each SSPP determination.
\label{fig:dd_all} }
\end{figure}

We also compare the shifts for the grid of pairs to those for the
primaries to see if there is any correlation between the two (see
Fig.\,\ref{fig:dd_all}). There appears to be small anticorrelations, all
less than a slope of $-$0.6, between the amount a primary is shifted and
the amount a pair is shifted, determined by performing linear least
squares fits on the points. These anticorrelations are small, indicating
that the addition of a secondary shifts the SSPP determinations
independently of the standard offsets. The largest anticorrelation is
$-$0.57 for the [$\alpha$/Fe] measurements.

For the following analyses, we examine the shifts in atmospheric
parameters due to an undetected secondary in two ways. First, we examine
the ``grid'' of synthetic spectra. Namely, we calculate the differences
between the synthetic binaries and their associated primaries
individually for every modeled spectra. Second, we examine the shifts of
a numerical population model of stars. The Galaxy does not have a flat
mass distribution for stars. To accurately determine the actual shifts
due to undetected binarity in our SEGUE G-K sample, we combine our
numerical population modeling with our grid of synthetic spectra
atmospheric parameters. We create a large sample including every
combination of primary and secondary mass distributions at every
metallicity as a model of the actual stellar sample. By matching each
primary and secondary in this sample with the parameters in the
synthetic spectra grid, we determine the most frequent shifts in
temperature, metallicity and surface gravity, in addition to the spread
in these shifts for a realization of the SEGUE G-K sample. We combine
the numerical sample of primaries with the grid of primaries vs. model
parameters to determine the uncertainties stemming from the
imperfections in our synthetic spectra and the SSPP's analysis of
them. This checks the SSPP measurements for a set of known inputs,
allowing us to calculate constant offsets associated with the
uncertainties in the modeling. We then determine the differences between
the sample of binaries and primaries. Analyzing the uncertainties from
the primaries with those for the binaries, we can isolate the
uncertainties that stem exclusively from binarity.

\subsection{Effective Temperature}
\label{sec:temp_shifts}
\begin{figure}
\centering \includegraphics[width=\textwidth]{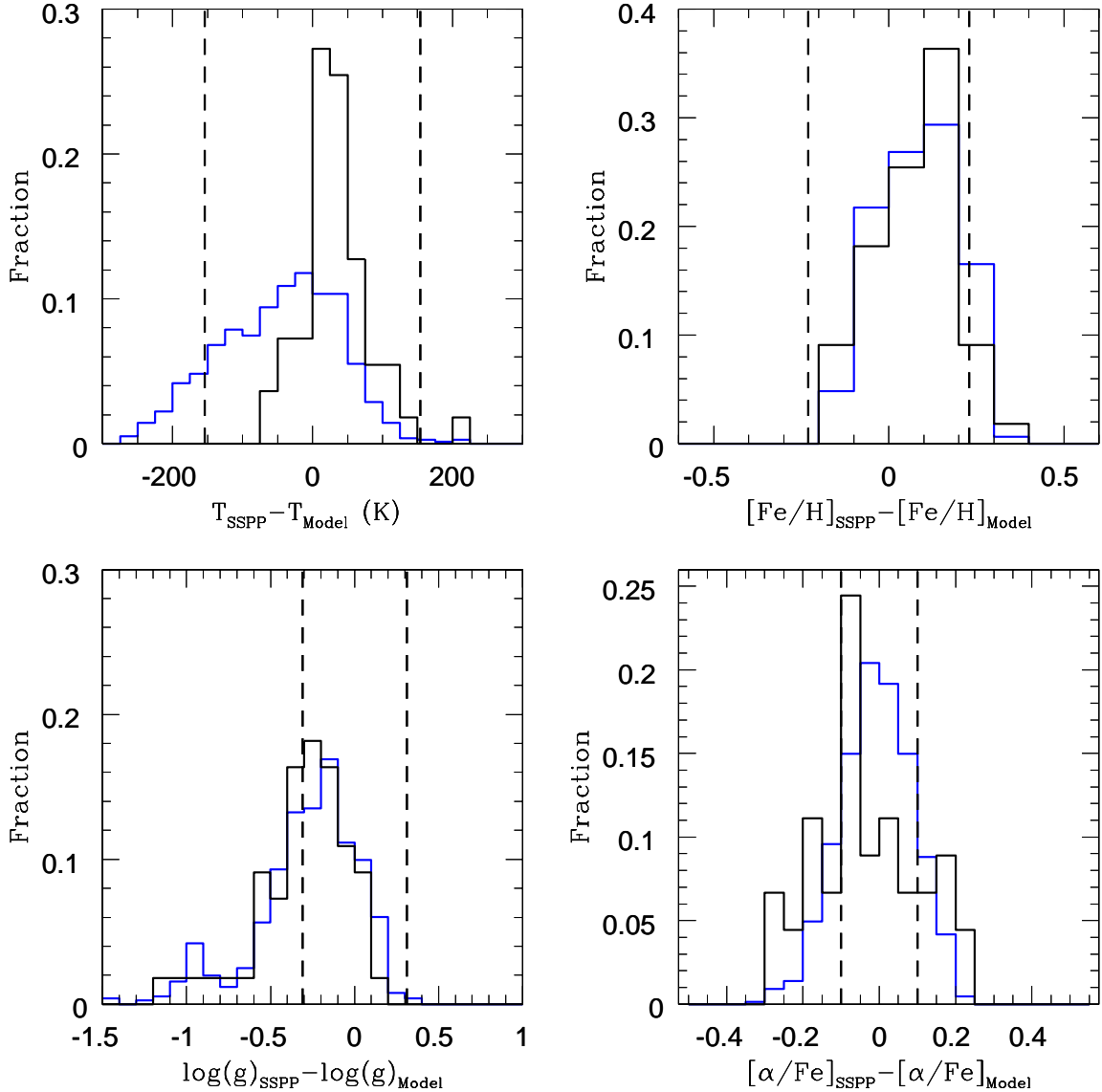}
\figcaption[ssppcomphist_all_new.eps]{The offsets between the pipeline
determinations for all of the modeled primaries and pairs for various
SSPP parameters.  The black histogram is the control group of primaries
and the blue is the blended binaries. The dashed lines represent the
SSPP errors for different determinations.  The top left figure is for
temperature, with a pipeline error of $\pm$150 K.  In general, the
offsets for both the primaries and pairs are well within the
uncertainties of the pipeline.  The top right is a comparison of [Fe/H]
determinations. Similarly, most of the shifts remain within the expected
errors of 0.24 dex. The bottom left is the surface gravity offsets, with
an error of 0.29 dex. Many of these are shifted out of the error range.
Lastly, the bottom right figure is for [$\alpha$/Fe] measurements using
errors of 0.1 dex (Lee et al. in prep).  For all but temperature the
shape of the offset histograms are quite similar for the pairs and
primaries, indicating that the addition of a secondary has little effect
on the SSPP determinations for metallicity, surface gravity, and alpha
enhancement.  As expected, the secondary does noticeably affect the
temperature determinations, systematically shifting them to cooler
values.
\label{fig:ssppcomp_hist_all} }
\end{figure}

The SSPP has temperature errors of 150 K at $S/N \geq$50
\citep{lee08_I}.  The addition of a secondary decreases the SSPP
measured temperature from that determined for the primary alone (see
Fig.\,\ref{fig:delta_all},\,\ref{fig:ssppcomp_hist_all}). This is
expected, as we can see by examining the changes in $(g-r)_0$ color (see
Fig.\,\ref{fig:gminr}). Whereas shifts in the other atmospheric
parameters may be due to random errors produced by the contribution of a
secondary, the downward shift in temperature is a systematic shift due
to the low mass secondaries being redder than the primaries.

The most frequent offset between the pair and the model temperature of
the primary for the grid is approximately -12.5 K, whereas it is
$\sim$12.5 K for the primary control group (see
Fig.\,\ref{fig:ssppcomp_hist_all}). The most extreme shift in the G-K
color range results from a 1\,\msun\, primary with [Fe/H]=$-$2.5 with a
0.85\,\msun\, secondary shifted in temperature by 410 K.  When we
examine our numerical population sample of primaries, we find that the
mode shift is 55 K with $\sigma(T)$ of 36 K for the largest $S/N$ (see
Table\,\ref{tab:shifts_and_sigmas}). This indicates the uncertainty from
imperfections in our synthetic spectra is $\pm$36 K.

\begin{figure}
\centering \includegraphics[width=\textwidth]{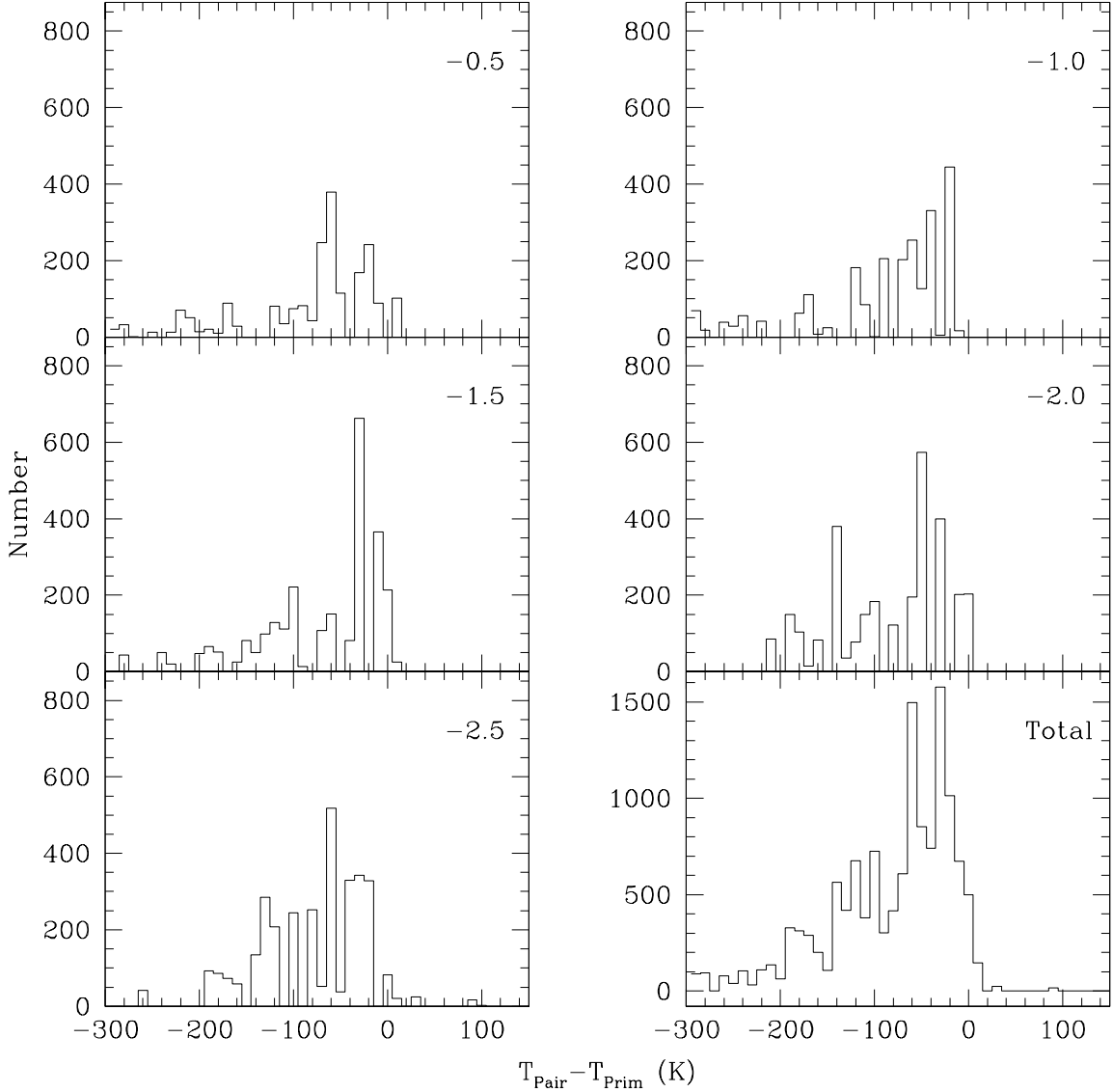}
\figcaption[teffdiff_updated2.eps]{The difference in SSPP-determined
temperature between the primaries and pairs for pairs within the G-K
color range.  The pair SSPP temperature is compared to the temperature
that SSPP calculates for the primary model spectrum alone. This isolates
the effect of the addition of a secondary, rather than discrepancies
resulting from the SSPP itself. Each plot is of a different metallicity
sample, with the combined sample on the bottom right. There is variation
with metallicity in the temperature differences. For the total 
sample, the mode is $\sim$15 K. The displayed figure is for a Chabrier
primary and secondary mass distribution, but for all combinations the
basic form of the histogram remains the same, as does the
mode. \label{fig:teffhists}}
\end{figure}

We then use our grid of synthetic spectra to examine the mode and
uncertainty in our binaries when compared to their associated
primaries. First, we examine the variation in the shifts at different
metallicities.  Fig.\,\ref{fig:teffhists} displays the sizes of the
shifts in temperature for all targets within the G-K color cut in a
10,000 target sample at each metallicity, and finally, for all of these
samples combined. There is some variation in the size and range of the
shifts at different metallicities. Additionally, there is some variation
in the shifts from different combinations of primary and secondary mass
functions. Table\,\ref{tab:temp_percent} lists the absolute value of the
shifts in temperature between binaries and primaries for spectra of
infinite $S/N$; the listed uncertainties reflect the variation in these
percentages from the different combinations of primary and secondary
mass functions. The uncertainties listed for the ``Total'' sample
account for variation in both mass functions and
metallicity. 52$\pm$10\% of the sample have the addition of a secondary
shift their temperature from that of the primary by $\leq$60 K, a
minimal effect (see Table\,\ref{tab:temp_percent}). In fact,
$\sim$82$\pm$7\% of the pairs are within 150 K of their associated
primary.  Only $\sim$18\% of the shifted pairs lie outside the SSPP
temperature uncertainties at the highest $S/N$.

\begin{figure}
\centering \includegraphics[width=\textwidth]{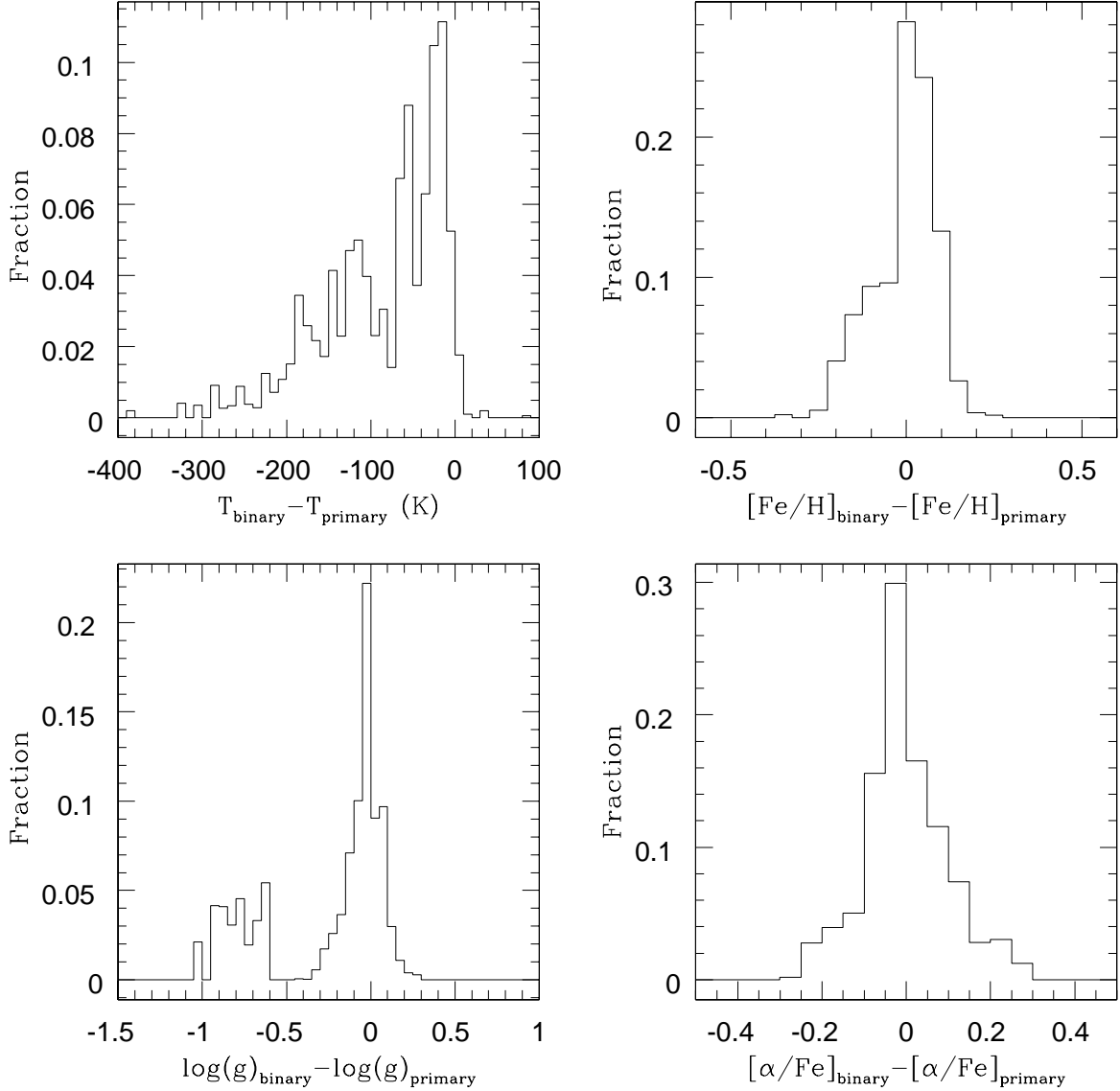}
\figcaption[binary_sni.eps]{The difference in the atmospheric parameters
determined by the SSPP for the binaries and primaries of a large
unbiased sample with infinite $S/N$. This sample includes every
combination of mass distributions at every metallicity. The percentage
of binaries shifted by certain amounts in temperature and metallicity
are listed in Table\,\ref{tab:temp_percent} and \ref{tab:feh_percent}.
There is some variation in the distributions with decreasing $S/N$ which
is reflected in Table\,\ref{tab:signaltonoise}, which lists the
confidence intervals, and Table\,\ref{tab:shifts_and_sigmas}, which
lists the most frequent shifts and spreads. \label{fig:binary_sni}}
\end{figure}

Our ``Total'' sample is our complete numerical population model,
including every combination of primary and secondary mass functions at
each metallicity. We use this sample to determine the effects of
binarity on our SEGUE sample. Examining the mode and $\sigma(T)$ shows
that the binary sample is most frequently shifted down 15 K in
temperature, with a variance of $\pm$72 K (see Fig.\,\ref{fig:binary_sni},
Table\,\ref{tab:shifts_and_sigmas}). This variance represents the
uncertainty from both binarity and the synthetic modeling errors, which
we measured above to be $\pm$36 K. We can isolate the uncertainty from
undetected secondaries using the values determined for the primaries
vs. the model parameters, assuming the errors add in quadrature. For
infinite $S/N$, there is a systematic shift of $-$15 K and an additional
uncertainty of $\pm$62 K due to undetected binaries.

\subsection{Metallicity}
The metallicity determinations are relatively unaffected by the addition
of a secondary.  The primary control group most often overestimates the
metallicity by $\sim$0.15 dex (see
Fig.\,\ref{fig:sspp_prim_feh}). Whereas the addition of a secondary
affected the shifts in temperature, there is little shift in the
metallicity determinations of pairs versus primaries (see
Fig.\,\ref{fig:ssppcomp_hist_all}). The largest metallicity shift for
targets in the $(g-r)_0$ color cut range defined for G and K dwarfs is
0.36 dex. This shift is for a pair of metallicity $-$0.5, a
0.75\,\msun\, primary with a 0.6\,\msun\, secondary.

Once again, we combine the shifts in the grid of synthetic spectra with
the numerical modeling to determine the metallicity uncertainties most
often found in a SEGUE sample of G-K dwarfs (see
Fig.\,\ref{fig:fehhists}).  Similar to the temperature shifts, there is
variation in the metallicity shift over the range of [Fe/H]. There is
little variation with different combinations of primary and secondary
mass distributions (see the uncertainties listed in
Table\,\ref{tab:feh_percent}). For the blended pairs with infinite
$S/N$, $\sim$62$\pm$3\% will be within 0.05 dex of the determination of
the primary; the addition of a secondary does not have a
significant effect (see Table\,\ref{tab:feh_percent}).  While $\sim$18\%
of the shifts in temperature are outside the SSPP uncertainties, only
$\sim$1\% of the binary sample are shifted by an amount greater than
$\sigma([Fe/H])$ from the SSPP for the infinite $S/N$ sample.

We expect the metallicity determinations of the pipeline to agree as
well if not better than temperature measurements, as both stars in the
pair are of the same metallicity. This is reflected in the mode and
$\sigma([Fe/H])$ determined for the large unbiased sample (see
Fig.\,\ref{fig:binary_sni} and Table\,\ref{tab:shifts_and_sigmas}).
Using the same methodology as in \S\,\ref{sec:temp_shifts}, we find that
a sample of G-K dwarf stars at infinite $S/N$ will have no systematic
shift but an additional uncertainty of around $\pm$0.05 dex.

\subsection{Surface Gravity}
We next consider the surface gravity determinations of the pipeline. The
control group of primaries is shifted down in surface gravity (see
Fig.\,\ref{fig:ssppcomp_prim_logg}), with a mode of $-$0.25 dex,
within the expected error of 0.29 dex \citep{lee08_II}. The addition of
a secondary does not significantly affect the surface gravity offsets
for the grid of synthetic spectra, as shown in
Fig.\,\ref{fig:ssppcomp_hist_all}, shifting the mode to 0.15 dex. 

Again, we apply the grid of synthetic values to our unbiased numerical
sample (see Fig.\,\ref{fig:binary_sni}). Using the same methods used for
temperature and metallicity, we find that the uncertainty from binarity
for the G-K dwarf sample is around $\pm$0.25 dex (see
Table\,\ref{tab:shifts_and_sigmas}). At different signal-to-noise
ratios, the uncertainties in the primary sample are similar or larger
than those of the binaries.  This implies that the effects of an
undetected secondary are minimal for measurements of $\log g$ in the
SSPP.

\subsection{[$\alpha$/Fe]}
Each of the models has a specified [$\alpha$/Fe] value defined by the
properties of the MARCS model atmospheres' metallicity and composition
model. As we are using the standard composition models, for
[Fe/H]=$-$0.5, [$\alpha$/Fe] is set to 0.20. For all our other
metallicities, [$\alpha$/Fe]=0.40. We compare the control group of
primaries to the [$\alpha$/Fe] determined by the SSPP in
Fig.\,\ref{fig:ssppcomp_prim_alpha}. The spread of measurements for the
grid of synthetic primaries is large, with a mode of -0.075 dex and a
range from approximately $-$0.3 to 0.3. The addition of a synthetic
secondary shifts the mode to -0.025 with a spread of shifts that looks
more Gaussian for the grid of binaries (see
Fig.\,\ref{fig:ssppcomp_hist_all}). A KS test indicates that the
distributions have a 60\% chance of being from the same parent sample.

Because we are spectroscopically blending pairs of the same metallicity
and composition (i.e. the primary and secondary have the same
[$\alpha$/Fe] values), we expect there to be little difference between
the parameters determined for the primaries and the blended binaries.
When we apply our grid of differences to the unbiased sample, we
calculate an uncertainty from binarity of around $\pm$0.10 dex (see
Table\,\ref{tab:shifts_and_sigmas}). However the uncertainties 
in the primary sample are comparable in size to those 
of the binaries, indicating that binarity is not a dominant 
uncertainty for measurements of [$\alpha$/Fe].

\subsection{Effects of the Signal-to-Noise Ratio}

It is possible that signal-to-noise ratio effects can diminish the
effect of a secondary. In particular, noise in the spectrum could
overwhelm any contributions from an undetected companion. We analyze
both the original infinite signal-to-noise synthetic spectra and also
degrade each model to a median signal-to-noise ($S/N$) of approximately
50, 25 and 10, covering the $S/N$ range of SEGUE's targets. A model of
the noise in SEGUE has been applied to high $S/N$ stellar spectra over a
range of spectral types to simulate spectra with signal-to-noise ratios
from 6 to 60 (C. Rockosi, private communication).  Using these
realizations, we calculate the $S/N$ at each point in the spectrum.  As
the noise patterns can potentially vary with spectral type, we match up
the SEGUE $S/N$ models to our synthetic spectra based on $(g-r)_0$
color.  We then compute a value for the noise fluctuation at each point
in our model spectra based on the $S/N$ of the noise-modeled spectra. We
convolve this noise value with a Gaussian, and add it to our spectra
signal-to-noise ratios of 50, 25, and 10.

\begin{figure}
\centering \includegraphics[width=\textwidth]{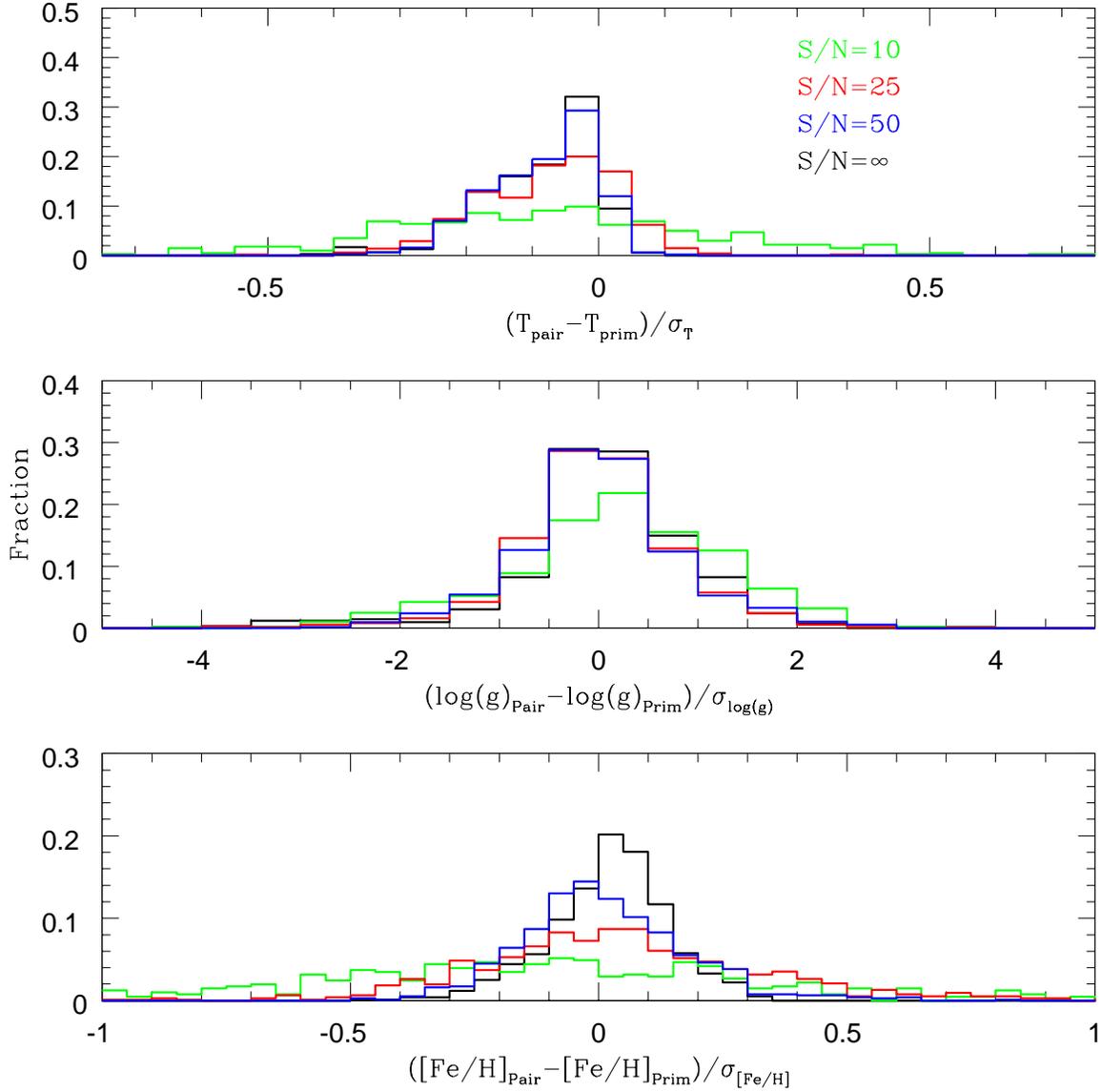}
\figcaption[sspp_noisecomp_scott2.mol.eps]{The effect of $S/N$ on the SSPP-determined 
atmospheric parameters. To isolate the difference degraded $S/N$ has on 
the SSPP measurements, we calculate the difference between the values determined for 
the pairs and the primaries at a given $S/N$ and divide it by the dispersion in 
that parameter determined by the SSPP for the primaries at that $S/N$ ($\sigma$). 
This compares the effect of 
binarity on a parameter at a given $S/N$ to the expected spread at that $S/N$. 
The top plot examines temperature, the middle surface 
gravity, and the bottom metallicity. The original infinite $S/N$ models 
are plotted in black, models with $S/N$ of 50 are in blue, $S/N$ of 25 
are in red, and lastly $S/N$ of 10 is plotted in green. \label{fig:noisy}}
\end{figure}

We run the degraded and infinite $S/N$ models of the control sample of
primaries and modeled binaries through the SSPP. To isolate the effect
an undetected secondary has on the parameter determination of the SSPP,
we compare the primary and pair values for each model at each $S/N$ (see
Fig.\,\ref{fig:noisy}). We determine the difference between the primary
and pair values and compare these to the variance of the parameter in the
control group, isolating the effect of a blended secondary at different
$S/N$.  As expected, as the $S/N$ decreases, the spread determined for
all atmospheric parameters increases. For $S/N>$10, the distributions
are quite similar, showing similar effects of binarity from a
signal-to-noise of $\sim$100 to 25.  For $S/N$ of 10, the spread of
values increases greatly. With spectra this noisy, the SSPP accuracy 
is already decreased significantly. With the spectral contribution of 
an undetected secondary, it is even more difficult for the SSPP to 
determine accurate atmospheric parameters. 

Combining our numerical and synthetic spectra modeling at a range of
$S/N$, we have determined the amount of shifting in temperature and
metallicity for the confidence intervals, 68\%, 95\%, and 99\% of our
modeled SEGUE sample (see Table\,\ref{tab:signaltonoise}).  Similar to
Fig.\,\ref{fig:noisy}, this table indicates that the distributions of
atmospheric parameter shifts are similar up until a $S/N$ of
10. Starting with the 1$\sigma$ interval, 68\% of the sample for the
entire range of $S/N$ is within $\sim$140 K, within the $S/N$=50
uncertainty of 150 K for the SSPP. Moving out to the 2$\sigma$ sample,
95\% of the modeled sample of $S/N=\infty$, 50, and 25 are within 230
K. However, 95\% of the modeled sample with $S/N=$10 are within 480 K, a
much greater number. This behavior is also seen for
metallicity. Similarly, we examined the mode and RMS of the shift for
the various median signal-to-noise ratios.  The uncertainties for the
various atmospheric parameters tend to increase as the signal-to-noise
decreases (see Table\,\ref{tab:shifts_and_sigmas}). This pattern
continues when we isolate the uncertainty of binarity alone from the
uncertainty from the synthetic spectra.

\section{Conclusions}

In this analysis, we have modeled samples of 100,000 binaries with
primaries from 0.5 to 1.0\,\msun\, and a variation of mass distributions
and metallicity, to better understand their effect as potential
contaminants of the SEGUE sample in the G-K dwarf range. Work by
\citet{dm91} established that around 65\% of F-G type stars have at
least one companion. Thus, understanding how undetected binaries affect
the atmospheric parameter determinations in SEGUE is crucial.
 
From our Monte Carlo analysis, we have determined that of a sample of
100,000 binaries, modeled using a range of mass distributions, on
average 90$\pm$1\% will appear spectroscopically or photometrically
blended in the SEGUE sample, i.e. the two stars will be within
3\arcsec\, of each other projected on the sky. Of all the pairs with G-K
type primaries, approximately 30$\pm$2\% of the sample of 100,000
binaries based on a $(g-r)_0$ color cut, $\sim$93\% are
blended. 

To quantify the effect of an undetected secondary on the stellar
atmospheric parameters T$_{eff}$, [Fe/H], and $\log g$ determined for
the SEGUE spectra, we utilized a grid of synthetic spectra processed by
the SSPP. We quantified the systematic offsets between the synthetic
spectra parameters and those measured by the SSPP, which result from the
various approximations made in our spectral modeling. We then compared
the determinations for the blended pairs to those of the primaries to
quantify the effect of an undetected companion. Examining the
distribution of offsets at infinite $S/N$ shows that the majority of the
G-K sample is within the established SSPP errors for both temperature
and metallicity (see Table\,\ref{tab:temp_percent},
\,\ref{tab:feh_percent}). In particular, 82$\pm$7\% of blended pairs
with a G-K dwarf primary with $S/N$ of $\sim$50 are within the SSPP's
error of 150 K in temperature. For determinations of [Fe/H], 99$\pm$1\%
of these pairs have a measured metallicity that differs from that of the
primary by less than the established SSPP metallicity error of 0.24 dex.
Examining the modeled pairs, we find that very few are outliers in both
temperature and metallicity. Of the 53 synthesized pair spectra in the
G-K color range that are outliers in temperature or metallicity, only 3
are outliers in both estimates.  Thus, we can assume that all outliers
in metallicity are independent of those in temperature, for a total of
$\sim$18$\pm$7\% of the G-K blended targets shifted a significant amount
in metallicity or temperature by an undetected secondary.

A search of SEGUE using CasJobs and based on the target selection
parameters extracted a data set of $\sim$20,000 G-K type dwarfs in the
sample.  According to statistics from \citet{dm91}, 13,000 of these are
in binaries. Applying our numerical results, we conservatively assume
that 93\% of these G-K binaries are blended pairs. Thus, of a sample of
20,000 G-K stars, $\sim$12,000, or 60\%, are potentially affected in
SEGUE by a secondary companion.  Using our spectroscopic analysis, we
can determine how many of this subsample are expected to have inaccurate
SSPP parameter determinations, due to their undetected companion.  We
determined that 18$\pm$1\% of the G-K blends are shifted beyond the
expected uncertainties in temperature and/or metallicity in the SSPP by
the presence of a secondary, a total of $\sim$2000 SEGUE targets. Thus,
11$\pm$2\% of the entire G-K dwarf sample of high signal-to-noise will
be significantly affected by an undetected companion in its SSPP
temperature or metallicity determination. This 11$\pm$2\% sample will be
systematically shifted to cooler temperatures, and generally shifted
down in metallicity as well. The percentage affected is similar for a
$S/N$ of 50.  For signal-to-noise of 25, the expected SSPP uncertainties
increase \citep{lee08_I}; $\sim$10\% are shifted outside the expected
uncertainties, similar to the value for $S/N$ of 50 and higher.  This
percentage increases significantly for $S/N$ of 10. $\sim$40\% of the
G-K dwarf sample will be shifted outside the expected uncertainties in
temperature and/or metallicity at this signal-to-noise.

Beyond examining the percentage of targets pushed beyond the SSPP
uncertainties in various atmospheric parameters, we quantify the
uncertainties from our synthetic spectra individually and from the
undetected secondary.  Both the systematic shift and additional spread
in each parameter must be taken into account when accounting for binary
contamination in the SEGUE sample.  The most frequent shift and spread
values we derive for each $S/N$ we model are summarized in
Table\,\ref{tab:shifts_and_sigmas}. For $\log g$ and [$\alpha$/Fe] the
most frequent shifts are very small. The uncertainties in these
parameters for the primary and binary samples are similar in
size. Sometimes the uncertainty measured for the primary sample is even
larger than that of the secondaries. The small shifts and variation in
the $\sigma$ indicates that, for these two parameters, the uncertainties
due to binarity are minimal with respect to the general uncertainties in
determining the values themselves. For temperature and metallicity
however, binarity can increase the SSPP uncertainties in a well defined
way, with it systematically decreasing the measured temperature and
slightly affecting the measured [Fe/H]. Additionally, the shifts in
metallicity are quite small, while there are clear systematic shifts
down in temperature.

An additional concern about binary contamination was its effect on
target selection, as SEGUE uses photometric color cuts to extract
different spectral types. Our analysis indicates that approximately 93\%
of all primaries that are within the G-K dwarf color cut,
0.48$\leq$$(g-r)_0$$\leq$0.75, remain within this cut with the addition
of a secondary (see Table\,\ref{tab:numbers}). The most frequent shift
is merely 0.01$\pm$0.02 in $(g-r)_0$. Thus, the target selection effect
of undetected binaries is small, but not entirely absent.

\begin{figure}
\centering \includegraphics[width=\textwidth]{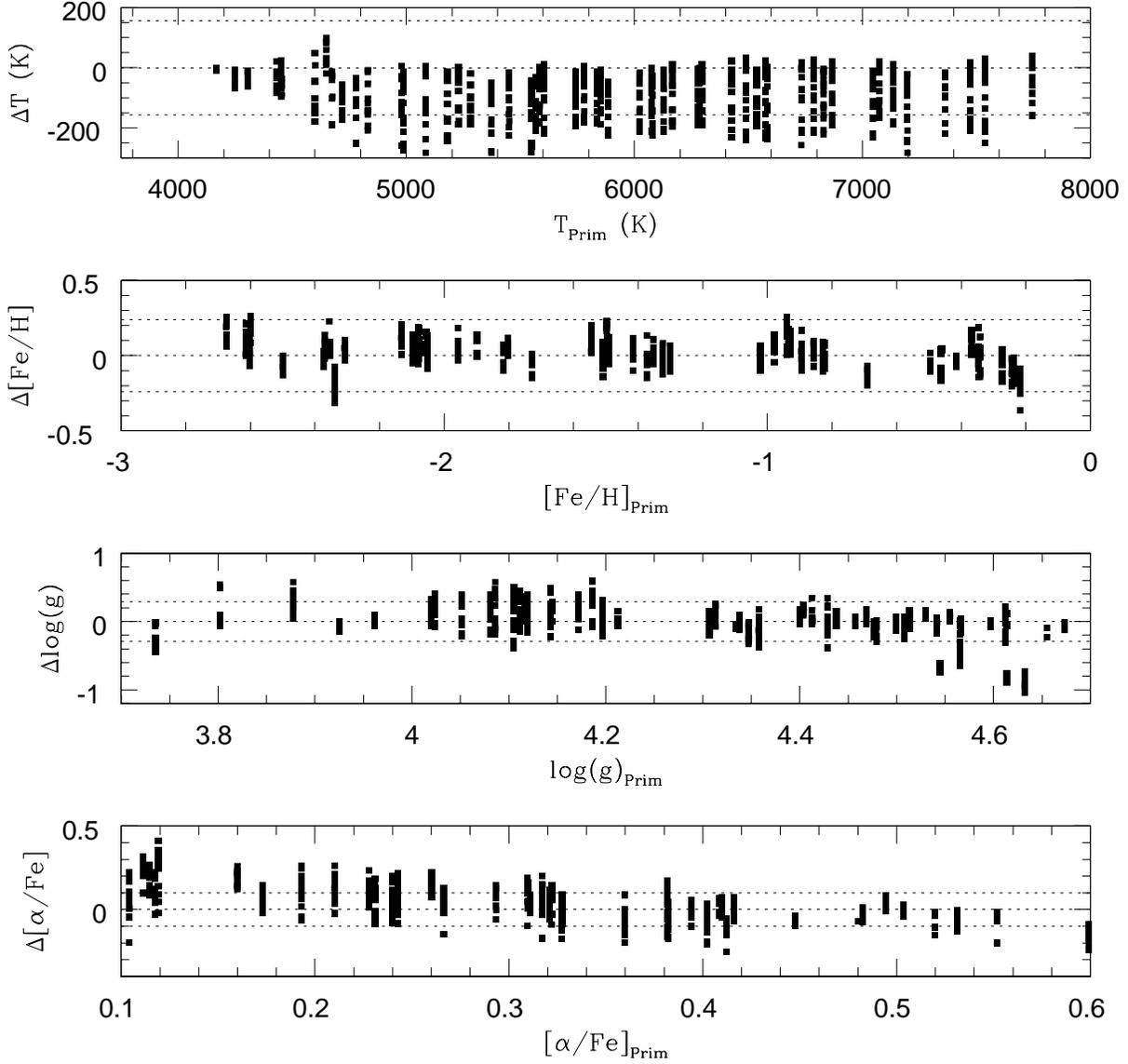}
\figcaption[delta_all3.eps]{A comparison between the SSPP output for the
control group of primaries and the binaries. Four parameters from the
pipeline are plotted: temperature, metallicity, surface gravity, and
alpha enhancement. The differences on the y axis are the output from the
pairs with the primary values subtracted. The x axis is the values for
the primaries. The dashed lines reflect the expected SSPP uncertainties 
for these calculations. These plots indicate that the shifts are not dependent on
the values for the primary and are, in general, quite small. The shifts
are consistent across the entire spread of values.
\label{fig:delta_all} }
\end{figure}

\begin{figure}
\centering \includegraphics[width=\textwidth]{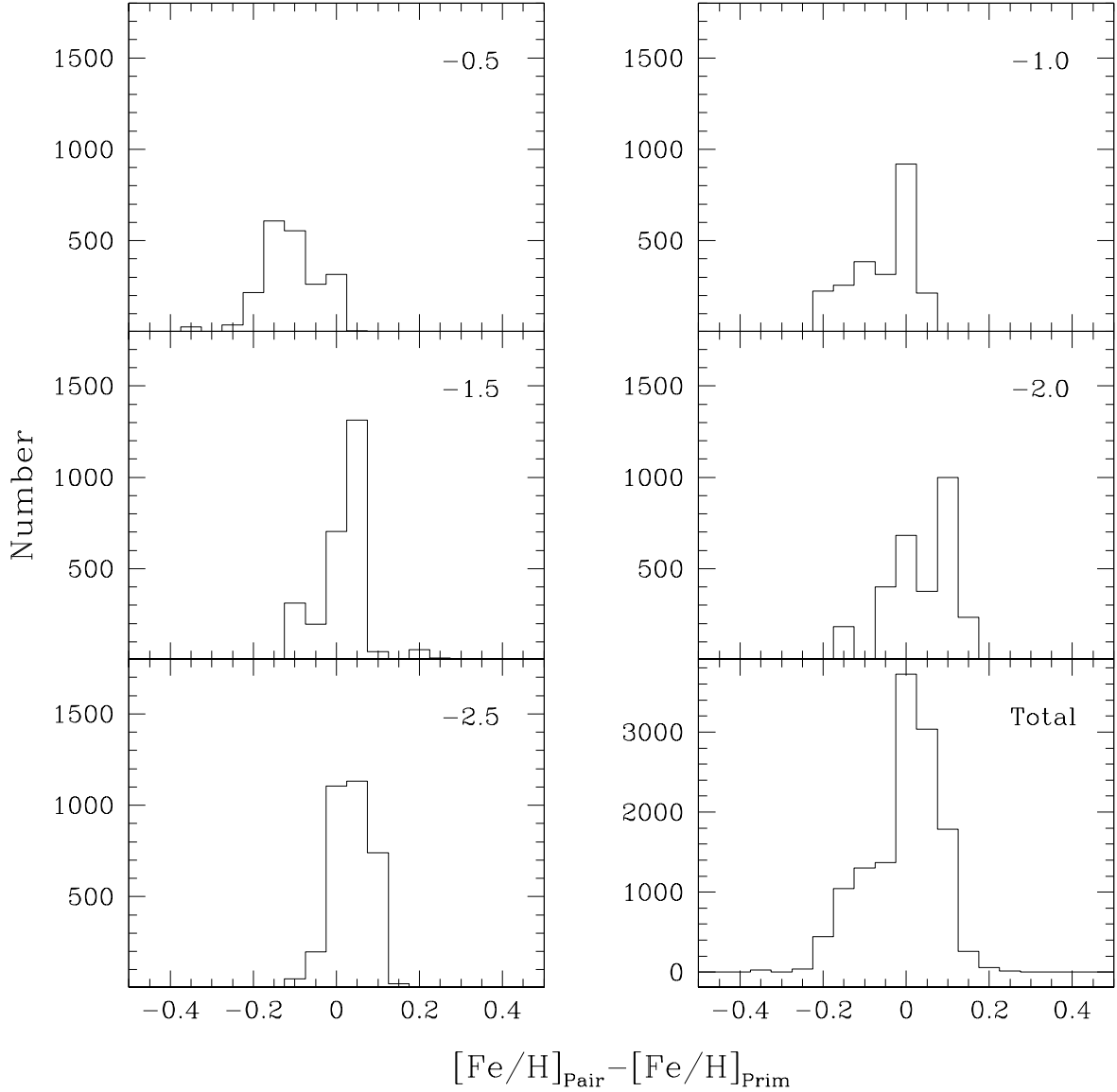}
\figcaption[fehdiff_updated2.eps]{Similar to Fig.\,\ref{fig:teffhists},
except for metallicity rather than temperature. The [Fe/H] determined
for the primaries by SSPP is compared to that measured for the pairs for
all pairs within the appropriate color range.  Like our temperature
analysis, the distribution of the difference varies with metallicity. In
general, the metallicities determined for the pairs agree quite well
with those of the primaries, with little to no shift, as expected.  The
mode for the shift of the total sample is $\sim$0.1 dex. As with the
temperature histogram, this figure is for a Chabrier primary and
secondary mass distribution; for all combinations, the basic form for
each histogram and the most frequent shift remain the
same. \label{fig:fehhists}}
\end{figure}

Finally, it is important to understand the effect of these undetected
binaries on the metallicity distribution function (MDF) of the Milky
Way. As noted earlier, due to their long lifetimes, G and K dwarfs are
valuable for understanding the early conditions of the Galaxy. Although
the shifts in metallicity are in general small over the entire range of
[Fe/H]=$-$0.5 to $-$2.5 for the modeled pairs (see
Fig.\,\ref{fig:delta_all}), when applied to the numerical models of
blended binaries in the G-K range, there is a tendency for lower
metallicity pairs to be shifted more in [Fe/H] (see
Fig.\,\ref{fig:fehhists}). Although the most frequent shift remains
small, there is increased spread in $\Delta$[Fe/H] with decreasing
metallicity. As it is more difficult to determine the metallicity for
low metallicity stars because their features are not as strong, we
expected there to be an increased spread at lower metallicity. This will
make the low-metallicity end of the MDF more uncertain. We can use 
our binary modeling to better understand the size of the binary contamination 
effect on the MDF. 

Our examination of the effects of undetected secondaries in the SEGUE
sample has established that for $S/N >$10, only around 10\% of G-K dwarf
type stars will have their derived atmospheric parameters (T$_{eff}$,
[Fe/H]) shifted by more than the SSPP errors at that signal-to-noise due
to an undetected companion. Additionally, the added uncertainties are
insignificant for $\log g$ and [$\alpha$/Fe].  Primarily, secondaries
serve to decrease the effective temperatures measured for the primary by
the SSPP, while the measurements og metallicity not significantly
altered, likely due to the fact that this value should be the same for
both members of the pair.

\acknowledgements 

K.S. and J.A.J acknowledge support from NSF grant AST-0807997.  Y.S.L
and T.C.B. acknowledge partial support from grant PHY 08-22648: Physics
Frontiers Center/Joint Institute for Nuclear Astrophysics (JINA),
awarded by the U.S. National Science Foundation. Funding for the SDSS
and SDSS-II has been provided by the Alfred P. Sloan Foundation, the
Participating Institutions, the National Science Foundation, the
U.S. Department of Energy, the National Aeronautics and Space
Administration, the Japanese Monbukagakusho, the Max Planck Society, and
the Higher Education Funding Council for England. The SDSS Web Site is
http://www.sdss.org/.

The SDSS is managed by the Astrophysical Research Consortium for the
Participating Institutions. The Participating Institutions are the
American Museum of Natural History, Astrophysical Institute Potsdam,
University of Basel, University of Cambridge, Case Western Reserve
University, University of Chicago, Drexel University, Fermilab, the
Institute for Advanced Study, the Japan Participation Group, Johns
Hopkins University, the Joint Institute for Nuclear Astrophysics, the
Kavli Institute for Particle Astrophysics and Cosmology, the Korean
Scientist Group, the Chinese Academy of Sciences (LAMOST), Los Alamos
National Laboratory, the Max-Planck-Institute for Astronomy (MPIA), the
Max-Planck-Institute for Astrophysics (MPA), New Mexico State
University, Ohio State University, University of Pittsburgh, University
of Portsmouth, Princeton University, the United States Naval
Observatory, and the University of Washington.

\clearpage

\include{table2}

\end{document}

%% file: table2.tex
\begin{deluxetable}{cccccc}
\tabletypesize{\small}
\tablewidth{0pt}
\tablecaption{Number and Percentage of blends in the G-K dwarf range \label{tab:numbers}}
\tablehead{
\colhead{Primary Distribution} & \colhead{Secondary Distribution} & \colhead{[Fe/H]} & \colhead{Primaries} & \colhead{Pairs} & \colhead{Percentage}}
\startdata
Salpeter & Salpeter  & -0.5 & 2566$\pm$ 7 & 1898$\pm$11 & 74$\pm$0.63 \\ 
         & Chabrier  & -0.5 & 2530$\pm$12 & 1886$\pm$11 & 75$\pm$0.75 \\ 
         & Kroupa    & -0.5 & 2526$\pm$ 9 & 1955$\pm$ 9 & 77$\pm$0.59 \\ 
         & DM        & -0.5 & 2538$\pm$11 & 1930$\pm$ 8 & 76$\pm$0.60 \\ 
         & Halbwachs & -0.5 & 2554$\pm$18 & 2000$\pm$13 & 78$\pm$0.96 \\ 
Kroupa   & Salpeter  & -0.5 & 2538$\pm$11 & 1873$\pm$ 9 & 74$\pm$0.66 \\ 
         & Chabrier  & -0.5 & 2526$\pm$11 & 1893$\pm$ 8 & 75$\pm$0.62 \\ 
         & Kroupa    & -0.5 & 2525$\pm$15 & 1974$\pm$17 & 78$\pm$1.07 \\
\enddata
\tablecomments{The numbers of primaries and pairs in the G-K color cut range for our models. 
A target must be within 0.48$\leq$$(g-r)_{O}$$\leq$0.75 to be selected as a G or K dwarf by SEGUE. 
In general, there are fewer photometrically blended pairs within the color cut than there 
are primaries, indicating that the addition of a secondary is more likely to bump a target 
out of G-K color range than in. Note, however, that for [Fe/H]=$-$1.0, there are more 
blended pairs in range than primaries. In general, 90$\pm$2\% of the primaries within 
the G-K dwarf color cut remain there when blended with a secondary, indicating that 
there is a noticeable, but not large, difference 
between the statistics with binary contamination. }
\end{deluxetable}

\begin{deluxetable}{cccccccccc}
\tabletypesize{\small}
\tablewidth{0pt}
\tablecaption{Parameters for the synthetic spectra modeled \label{tab:parameters}}
\tablehead{
\colhead{M (\msun)} & \colhead{[Fe/H]} & \colhead{T (K)} & \colhead{$\log g$} & \colhead{$\log (L/L_{\odot})$} & \colhead{u} & \colhead{g} & \colhead{r} & \colhead{i} & \colhead{z} }
\startdata
0.10 & -0.5 & 3275 & 5.35 & -2.89 & 18.40 & 14.80 & 13.20 & 12.15 & 11.60 \\ 	   
0.15 & -0.5 & 3325 & 5.20 & -2.51 & 17.10 & 13.75 & 12.20 & 11.20 & 10.70 \\ 	   
0.20 & -0.5 & 3400 & 5.10 & -2.26 & 16.20 & 13.00 & 11.40 & 10.50 & 10.10 \\ 	   
0.25 & -0.5 & 3500 & 5.00 & -2.07 & 15.40 & 12.30 & 10.80 & 10.00 & 9.60 \\ 	   
0.30 & -0.5 & 3550 & 5.00 & -1.92 & 14.90 & 11.80 & 10.35 & 9.60 & 9.20 \\ 	   
0.35 & -0.5 & 3625 & 4.90 & -1.77 & 14.30 & 11.35 & 9.90 & 9.20 & 8.80 \\ 	   
0.40 & -0.5 & 3700 & 4.90 & -1.62 & 13.80 & 10.90 & 9.45 & 8.80 & 8.50 \\ 	   
0.45 & -0.5 & 3800 & 4.80 & -1.46 & 13.20 & 10.30 & 8.95 & 8.30 & 8.10 \\
\enddata
\tablecomments{Parameters for the synthetic spectra modeled with 
MARCS model atmospheres in conjunction with TurboSpectrum. These 
values are determined from the Dartmouth isochrones \citep{dartmouth}.}
\end{deluxetable}

\begin{landscape}
\begin{deluxetable}{ccccccccccccccc}
\tabletypesize{\tiny}
\tablewidth{0pt}
\tablecaption{Properties of the control group of primaries analyzed by the SSPP \label{tab:control_group}}
\tablehead{
\colhead{M$_{Prim}$} & \colhead{Model T$_{eff}$} & \colhead{SSPP T$_{eff}$} & \colhead{Model [Fe/H]} & \colhead{SSPP [Fe/H]} & \colhead{Model $\log g$} & \colhead{SSPP $\log g$} & \colhead{Model [$\alpha$/Fe]} & \colhead{SSPP [$\alpha$/Fe]} & \colhead{u} & \colhead{g} & \colhead{r} & \colhead{i} & \colhead{z} & \colhead{(g-r)}}
\startdata
0.50 & 3950 & 4166$\pm$28 & -0.5 & -0.46$\pm$0.03 & 4.8 & 4.65$\pm$0.02 & 0.2 & - & 12.56 & 9.73 & 8.4 & 7.87 & 7.66 & 1.33 \\ 
0.55 & 4150 & 4249$\pm$50 & -0.5 & -0.41$\pm$0.09 & 4.7 & 4.67$\pm$0.17 & 0.2 & - & 11.85 & 9.07 & 7.84 & 7.39 & 7.24 & 1.23 \\ 
0.60 & 4425 & 4453$\pm$48 & -0.5 & -0.37$\pm$0.09 & 4.7 & 4.78$\pm$0.25 & 0.2 & - & 11.04 & 8.33 & 7.26 & 6.91 & 6.82 & 1.07 \\ 
0.65 & 4750 & 4721$\pm$67 & -0.5 & -0.50$\pm$0.01 & 4.7 & 4.51$\pm$0.09 & 0.2 & 0.38$\pm$0.01 & 9.96 & 7.58 & 6.71 & 6.44 & 6.4 & 0.87 \\ 
0.70 & 5050 & 4989$\pm$36 & -0.5 & -0.46$\pm$0.03 & 4.7 & 4.48$\pm$0.05 & 0.2 & 0.38$\pm$0.04 & 8.96 & 6.94 & 6.23 & 6.02 & 6.01 & 0.71 \\ 
0.75 & 5350 & 5374$\pm$69 & -0.5 & -0.22$\pm$0.01 & 4.6 & 4.48$\pm$0.06 & 0.2 & 0.45$\pm$0.03 & 8.09 & 6.4 & 5.81 & 5.64 & 5.66 & 0.59 \\ 
0.80 & 5600 & 5604$\pm$45 & -0.5 & -0.24$\pm$0.03 & 4.6 & 4.46$\pm$0.07 & 0.2 & 0.41$\pm$0.01 & 7.36 & 5.93 & 5.43 & 5.29 & 5.34 & 0.50 \\ 
0.85 & 5850 & 5884$\pm$76 & -0.5 & -0.24$\pm$0.01 & 4.6 & 4.44$\pm$0.08 & 0.2 & 0.23$\pm$0.03 & 6.78 & 5.51 & 5.09 & 4.98 & 5.05 & 0.42 \\
\enddata
\tablecomments{The SSPP parameter determinations for the control group as compared to the actual synthetic model values. Note that there are no 
temperature or metallicity measurements for stars less than 4000 K, as the SSPP does not work for 
these temperatures. The parameters for spectral modeling are listed as well. }
\end{deluxetable}
\end{landscape}

\begin{landscape}
\begin{deluxetable}{ccccccccccccc}
\tabletypesize{\small}
\tablewidth{0pt}
\tablecaption{Properties of the blended binaries analyzed by the SSPP \label{tab:binary_params}}
\tablehead{
\colhead{M$_{Prim}$} & \colhead{M$_{Sec}$} & \colhead{[Fe/H]$_{model}$} & \colhead{T$_{eff}$} & \colhead{[Fe/H]} & \colhead{$\log g$} & \colhead{[$\alpha$/Fe]}
& \colhead{u} & \colhead{g} & \colhead{r} & \colhead{i} & \colhead{z} & \colhead{(g-r)}}
\startdata
0.50 & 0.10 & -0.5 & - & - & - & - & 12.56 & 9.72 & 8.39 & 7.85 & 7.63 & 1.33 \\ 
0.50 & 0.15 & -0.5 & - & - & - & - & 12.55 & 9.70 & 8.37 & 7.82 & 7.60 & 1.33 \\ 
0.50 & 0.20 & -0.5 & - & - & - & - & 12.52 & 9.68 & 8.34 & 7.78 & 7.55 & 1.34 \\ 
0.50 & 0.25 & -0.5 & - & - & - & - & 12.49 & 9.64 & 8.29 & 7.72 & 7.49 & 1.35 \\ 
0.50 & 0.30 & -0.5 & - & - & - & - & 12.44 & 9.58 & 8.24 & 7.66 & 7.42 & 1.34 \\ 
0.50 & 0.35 & -0.5 & - & - & - & - & 12.37 & 9.51 & 8.16 & 7.58 & 7.34 & 1.35 \\ 
0.50 & 0.40 & -0.5 & - & - & - & - & 12.26 & 9.41 & 8.05 & 7.47 & 7.24 & 1.36 \\ 
0.50 & 0.45 & -0.5 & - & - & - & - & 12.09 & 9.24 & 7.89 & 7.32 & 7.10 & 1.35 \\
\enddata
\tablecomments{The temperature and metallicity determinations from SSPP for the pairs in addition 
to new photometric values. Note that there are no 
temperature or metallicity measurements for stars less than 4000 K. This is a small section of the 
total data table (i.e. just for [Fe/H]$_{model}$=$-0.5$) which is available on-line at http://www.astronomy.ohio-state.edu/$\sim$kschles/secondaries.pdf}
\end{deluxetable}
\end{landscape}

\begin{landscape}
\begin{deluxetable}{ccccccccccccc}
\tabletypesize{\small}
\tablewidth{0pt}
\tablecaption{Differences between the blended binaries and control group primaries \label{tab:binary_diff}}
\tablehead{
\colhead{M$_{Prim}$} & \colhead{M$_{Sec}$} & \colhead{[Fe/H]$_{model}$} & \colhead{$\Delta$T$_{eff}$} & \colhead{$\Delta$[Fe/H]} & \colhead{$\Delta \log g$} & \colhead{$\Delta$[$\alpha$/Fe]} & \colhead{$\Delta$u} & \colhead{$\Delta$g} & \colhead{$\Delta$r} & \colhead{$\Delta$i} & \colhead{$\Delta$z} & \colhead{$\Delta$(g-r)} }
\startdata
0.50 & 0.10 & -0.5 & - & - & - & - & 0.00 & 0.01 & 0.01 & 0.02 & 0.03 & 0.00 \\ 
0.50 & 0.15 & -0.5 & - & - & - & - & 0.01 & 0.03 & 0.03 & 0.05 & 0.06 & 0.00 \\ 
0.50 & 0.20 & -0.5 & - & - & - & - & 0.04 & 0.05 & 0.06 & 0.09 & 0.11 & -0.01 \\ 
0.50 & 0.25 & -0.5 & - & - & - & - & 0.07 & 0.09 & 0.11 & 0.15 & 0.17 & -0.02 \\ 
0.50 & 0.30 & -0.5 & - & - & - & - & 0.12 & 0.15 & 0.16 & 0.21 & 0.24 & -0.01 \\ 
0.50 & 0.35 & -0.5 & - & - & - & - & 0.19 & 0.22 & 0.24 & 0.29 & 0.32 & -0.02 \\ 
0.50 & 0.40 & -0.5 & - & - & - & - & 0.30 & 0.32 & 0.35 & 0.40 & 0.42 & -0.03 \\ 
0.50 & 0.45 & -0.5 & - & - & - & - & 0.47 & 0.49 & 0.51 & 0.55 & 0.56 & -0.02 \\
\enddata
\tablecomments{Change in parameters determined by the SSPP for the binaries with respect to the control group of primaries. The 
differences calculated are the values for the binaries with the primary values subtracted. Note again that there are no 
temperature or metallicity measurements for stars less than 4000 K. This is a small section of the 
total data table which is available on-line at http://www.astronomy.ohio-state.edu/$\sim$kschles/differences.pdf}
\end{deluxetable}
\end{landscape}

\begin{deluxetable}{ccccccc}
\tabletypesize{\small}
\tablewidth{0pt}
\tablecaption{Percent of blended G-K sample that is shifted in temperature by certain amounts \label{tab:temp_percent}}
\tablehead{
\colhead{Shift (K)} & \colhead{-0.5} & \colhead{-1.0} & \colhead{-1.5} & \colhead{-2.0} & \colhead{-2.5} & \colhead{Total} }
\startdata
0 & 0.00$\pm$0.00 &  0.00$\pm$0.00 &  10.15$\pm$3.35 &  10.86$\pm$2.10 &  2.76$\pm$0.84 &  5.19$\pm$1.29 \\ 
$\leq$10 & 9.68$\pm$2.03 &  1.56$\pm$1.61 &  22.69$\pm$3.85 &  20.80$\pm$3.68 &  5.01$\pm$1.67 &  12.23$\pm$1.56 \\ 
$\leq$20 & 18.26$\pm$6.00 &  19.58$\pm$5.67 &  22.69$\pm$3.85 &  20.80$\pm$3.68 &  16.99$\pm$2.33 &  19.63$\pm$3.25 \\ 
$\leq$30 & 27.04$\pm$7.34 &  20.68$\pm$5.05 &  47.67$\pm$9.79 &  32.60$\pm$5.68 &  30.00$\pm$4.93 &  32.10$\pm$5.92 \\ 
$\leq$40 & 27.04$\pm$7.34 &  33.97$\pm$6.92 &  50.75$\pm$9.73 &  32.60$\pm$5.68 &  38.59$\pm$6.30 &  37.19$\pm$6.87 \\ 
$\leq$50 & 32.41$\pm$7.37 &  40.26$\pm$8.34 &  50.75$\pm$9.73 &  47.05$\pm$8.74 &  39.73$\pm$6.28 &  42.61$\pm$7.84 \\ 
$\leq$60 & 50.18$\pm$13.35 &  47.51$\pm$10.55 &  55.67$\pm$10.75 &  52.26$\pm$10.02 &  52.10$\pm$9.16 &  51.77$\pm$10.45 \\ 
$\leq$70 & 59.82$\pm$14.61 &  54.68$\pm$12.20 &  59.24$\pm$11.06 &  52.26$\pm$10.02 &  53.20$\pm$9.06 &  55.44$\pm$10.92 \\
\enddata
\tablecomments{The percent of blended G-K sample that is shifted by a specific amount in SSPP temperature from 
the control group primary value. For example, $\approx$65\% 
of pairs are within 100 K of the pipeline temperature of their primary. The majority of these pairs are within 
60 K of the SSPP temperature determined for the primary alone. The uncertainties listed 
reflect the spread in percentages resulting from different combinations of primary and secondary mass distributions. 
The values determined for the ``Total'' sample reflect the combination of shifts for all metallicities. }
\end{deluxetable}

\begin{deluxetable}{ccccccccc}
\tabletypesize{\small}
\tablewidth{0pt}
\tablecaption{The systematic shifts and 1$\sigma$ uncertainties determined for both 
a sample of control group primaries and synthesized binaries. \label{tab:shifts_and_sigmas}}
\tablehead{\colhead{$S/N$} & \multicolumn{2}{c}{$\Delta$T (K)} & \multicolumn{2}{c}{$\Delta$[Fe/H] (dex)} & \multicolumn{2}{c}{$\Delta$log(g) (dex)} & \multicolumn{2}{c}{$\Delta$[$\alpha$/Fe] (dex)}  \\
   & Mode & $\sigma$ & Mode & $\sigma$ & Mode & $\sigma$ & Mode & $\sigma$ }
\startdata
\multicolumn{9}{c}{Primaries vs. Model Parameters} \\
$\infty$  & 55  & 36  & 0.20 & 0.08 & -0.18 & 0.26 & 0.18   & 0.13 \\
50        & 65  & 42  & 0.20 & 0.10 & -0.93 & 0.33 & 0.08   & 0.13 \\
25        & -75 & 71  & 0.35 & 0.12 & -0.23 & 0.41 & 0.08   & 0.13 \\
10        & -15 & 81  & 0.35 & 0.28 & -1.23 & 0.48 & 0.03   & 0.12 \\
\multicolumn{9}{c}{Binaries vs. Primaries} \\
$\infty$  & -15 & 72  &  0.0  & 0.09 & -0.03 & 0.36 & 0.03  & 0.10 \\
50        & -75 & 68  & -0.05 & 0.11 & -0.03 & 0.16 & -0.08 & 0.10 \\
\enddata 
\tablecomments{
The most frequent shift and the spread of differences between the synthesized models and 
the control groups at a range of signal-to-noise for a large numerically unbiased sample. 
The top section is the mode shift 
in the atmospheric parameters determined for the control group of primaries sample by the 
SSPP and the model parameters. Also listed is the 1$\sigma$ uncertainty associated 
with these differences. 
The next section lists the difference between the modeled sample binaries and control group of 
primaries. Whereas the top section reflects shifts and uncertainties associated with 
our spectral modeling and the SSPP, the middle section lists the uncertainties from 
undetected binaries. These shifts and uncertainties are determined from an initial 
sample of 750,000 targets, which includes every combination of primary and secondary 
mass functions at every metallicity. 
The following section is 
the uncertainties for these parameters determined for the SSPP. The uncertainties resulting from 
our modeling and the uncertainties from undetected binaries for our grid, in particular 
for temperature and metallicity, tend to be within the 
uncertainties expected for the SSPP \citep{lee08_I}. 
Finally, the bottom section is the isolated uncertainties from undetected binarity, determined 
from the errors of the control group and the binaries. The uncertainties for $\log g$ and 
[$\alpha$/Fe] how a lot of variation with $S/N$ and there is no clear increase in 
uncertainty with the addition of a secondary. This indicates that the effect of binarity 
is far from a dominant error in determining these parameters. }
\end{deluxetable}

\begin{deluxetable}{ccccccc}
\tabletypesize{\small}
\tablewidth{0pt}
\tablecaption{Percent of blended G-K sample that is shifted in metallicity by certain amounts \label{tab:feh_percent}}
\tablehead{
\colhead{Shift (dex)} & \colhead{-0.5} & \colhead{-1.0} & \colhead{-1.5} & \colhead{-2.0} & \colhead{-2.5} & \colhead{Total} }
\startdata
0 & 13.26$\pm$4.17 & 38.88$\pm$5.91 & 28.86$\pm$3.22 & 25.18$\pm$3.16 & 32.22$\pm$3.39 & 28.29$\pm$0.78 \\ 
$\leq$0.05 & 27.38$\pm$3.89 & 61.13$\pm$4.98 & 84.45$\pm$4.97 & 52.31$\pm$2.78 & 74.00$\pm$4.78 & 62.08$\pm$2.79 \\ 
$\leq$0.10 & 55.05$\pm$4.98 & 77.33$\pm$2.65 & 97.72$\pm$0.48 & 83.81$\pm$2.30 & 97.75$\pm$1.84 & 84.77$\pm$1.51 \\ 
$\leq$0.15 & 82.60$\pm$7.25 & 87.70$\pm$3.28 & 97.76$\pm$0.49 & 99.92$\pm$0.07 & 99.79$\pm$0.13 & 94.79$\pm$1.49 \\ 
$\leq$0.20 & 95.18$\pm$3.19 & 100.00$\pm$0.00 & 99.36$\pm$0.18 & 100.00$\pm$0.00 & 99.79$\pm$0.13 & 99.11$\pm$0.43 \\ 
$\leq$0.25 & 98.54$\pm$1.13 & 100.00$\pm$0.00 & 100.00$\pm$0.00 & 100.00$\pm$0.00 & 100.00$\pm$0.00 & 99.78$\pm$0.16 \\ 
$\leq$0.30 & 98.54$\pm$1.13 & 100.00$\pm$0.00 & 100.00$\pm$0.00 & 100.00$\pm$0.00 & 100.00$\pm$0.00 & 99.78$\pm$0.16 \\ 
$\leq$0.35 & 100.00$\pm$0.00 & 100.00$\pm$0.00 & 100.00$\pm$0.00 & 100.00$\pm$0.00 & 100.00$\pm$0.00 & 100.00$\pm$0.00 \\ 
$\leq$0.40 & 100.00$\pm$0.00 & 100.00$\pm$0.00 & 100.00$\pm$0.00 & 100.00$\pm$0.00 & 100.00$\pm$0.00 & 100.00$\pm$0.00 \\ 
\enddata
\tablecomments{The percent of blended G-K sample that is shifted in SSPP metallicity from the control group by different values in dex. Around 95\% of 
the sample are within 0.15 dex of their primary determinations, indicating that most shifts are within the errors. The uncertainties listed 
reflect the spread in percentages resulting from different combinations of primary and secondary mass distributions. The column 
named ``Total'' is the shift for all metallicities. }
\end{deluxetable}

\begin{deluxetable}{ccccccc}
\tabletypesize{\small}
\tablewidth{0pt}
\tablecaption{Statisics on the distribution of differences between modeled primaries and binaries at a range of $S/N$  \label{tab:signaltonoise}}
\tablehead{
\colhead{$S/N$} & & \colhead{Temperature} & & & \colhead{Fe/H} &   \\
              & 68\% & 95\% & 99\% & 68\% & 95\% & 99\% }
\startdata
$\infty$         & 110 K & 215 K & 280 K & 0.08 & 0.16 & 0.21 \\ 
50               & 110 K & 205 K & 280 K & 0.09 & 0.24 & 0.29 \\
25               & 113 K & 233 K & 323 K & 0.14 & 0.31 & 0.43 \\
10               & 136 K & 480 K & 495 K & 0.35 & 0.65 & 0.73 \\
\enddata
\tablecomments{The shift in temperature and metallicity for our samples at a range 
of signal-to-noise based on intervals, i.e. 68\% of synthetic binaries with $S/N$$=$25 are 
within 113 K of the temperature determined for their associated binary. The values for the 
$S/N$$=$50 are quite similar to our original infinite $S/N$ spectra; additionally, both are within the 
uncertainties of the SSPP in [Fe/H] to 95\%. The $S/N$$=$10 models have a significantly larger shifts 
in both temperature and metallicity, as expected (see Fig.\,\ref{fig:noisy}).}
\end{deluxetable}